\documentclass[conference]{IEEEtran}
\IEEEoverridecommandlockouts
\usepackage{cite}
\usepackage{amsmath,amssymb,amsfonts}
\usepackage{algorithmic}
\usepackage{graphicx}
\usepackage{multirow}
\usepackage{textcomp}
\usepackage{etoolbox}
\usepackage{subcaption}
\usepackage{cite}
\makeatletter
\patchcmd{\@makecaption}
  {\scshape}
  {}
  {}
  {}
\makeatother
\def\BibTeX{{\rm B\kern-.05em{\sc i\kern-.025em b}\kern-.08em
    T\kern-.1667em\lower.7ex\hbox{E}\kern-.125emX}}
\begin{document}

\title{A cow structural model for video analytics of cow health}
\author{\IEEEauthorblockN{He Liu, Amy R. Reibman, Jacquelyn P. Boerman}
\IEEEauthorblockA{Purdue University, 501 Northwestern Ave, West Lafayette, IN 47907, USA}
}

\maketitle

\begin{abstract}

In livestock farming, animal health directly influences productivity.
For dairy cows, many health conditions can be evaluated by trained observers based on visual appearance and movement.
However, to manually evaluate every cow in a commercial farm is expensive and impractical.
This paper introduces a video-analytic system which automatically detects the cow structure from captured video sequences.
A side-view cow structural model is designed to describe the spatial positions of the joints (keypoints) of the cow, and we develop a system using deep learning to automatically extract the structural model from videos.
The proposed detection system can detect multiple cows in the same frame and provide robust performance under practical challenges like obstacles (fences) and poor illumination.
Compared to other object detection methods, this system provides better detection results and successfully isolates the keypoints of each cow even when they are close to each other.


\end{abstract}


\section{Introduction} \label{sec:intro}

Monitoring animal health is a critical component of livestock farming, because healthy animals are more productive. 
Such monitoring is often performed visually, because animal appearance and behavior are key indicators of health changes. 
For example, trained farm personnel can analyze a dairy cow's health condition based on visual appearance \cite{fleishman2000some}, and can detect potential illnesses such as lameness \cite{cook2009influence}.
However, time and labor limitations preclude a human routinely watching for these changes, especially in commercial farms which house a large number of cows.
Thus there is increasing interest in substituting automated video analytics for human observations.
Indeed, video analytic techniques have been applied to different industries including animal agriculture.
With the help of the latest computer vision and image processing algorithms, visual animal biometrics has become an emerging research topic \cite{kumar2016visual}.
By applying video analytics methods, it is possible to develop a camera system that automatically detects the cow's health condition with low cost.


The first step to analyze cows using visual data is to detect and segment the cows within the video sequences.
This is a straightforward task when each cow walks individually on a well-lit pathway with a very clear background and no obstructions. 
However, if a camera system is installed on a commercial farm, with the goal of not interrupting daily operations, unrelated objects such as humans and fences are often captured.
So identifying the spatial locations of cows is a fundamental first step for further analysis.
However, finding the location of the cow is not enough. 
For the purpose of assessing an aspect of the animal (i.e. body size or gait), simply having a binary mask that labels the cow's location is inadequate.
Further information of the cow's structure is required, such as the locations of all body parts or joints \footnote{In this paper, we use the term body parts or keypoints to refer to the points that represent joints or specific regions on the cow.}.
This information can then be converted to human interpretable knowledge, or further processed by autonomous health monitoring systems.
In summary, we need a video system that not only isolates the cow's spatial location, but also detects its body structure and tracks its movements.

Designing a video processing system that satisfies these two requirements is not trivial.
Many previous segmentation methods focus on object detection, which generates a binary mask of the objects and their corresponding labels.
But this is not enough for further cow health analysis.
Recently, additional methods have been proposed to detect keypoint-based object structures, like human skeleton models.
These models are formed with a series of keypoints or joints connected in a particular order.
But these methods are designed with the knowledge of human structure, which is difficult to adapt to other animals like cows.
While new methods such as DeepLabCut \cite{mathis2018deeplabcut} focus on animal-related keypoint detection methods,
this method requires clear video sequences with a single object and a clear background. 
It cannot be directly applied for practical cow applications on the farm.
Finally, there are some visual applications \cite{pluk2010evaluation, zhao2018automatic, poursaberi2010real} for cows in the literature, but they are designed for videos captured in a specifically-designed environment, which requires extra efforts and costs for the farm to collect video data.

Processing cow videos collected from a commercial farm also poses specific challenges.
First, the environment in which video is captured cannot be fully controlled without interrupting the daily operation of the farm.
The cameras need to be installed with specific positions and viewing angles, so that the cows can be clearly observed with few obstructions.
Issues such as poor illumination \cite{zhao2018automatic} and heavy obstacles \cite{ter2017bootstrapping} largely influence the performance of existing detection algorithms.
In addition, the environment also limits the choice of capturing devices.
Surveillance cameras are the most suitable devices to install and deploy on typical farms, but the quality of the videos is limited.
Distortions such as low color saturation, low frame rate, and heavy compression deteriorate the performance of detection algorithms.

\begin{figure}
    \centering
    \includegraphics[width=8.8cm]{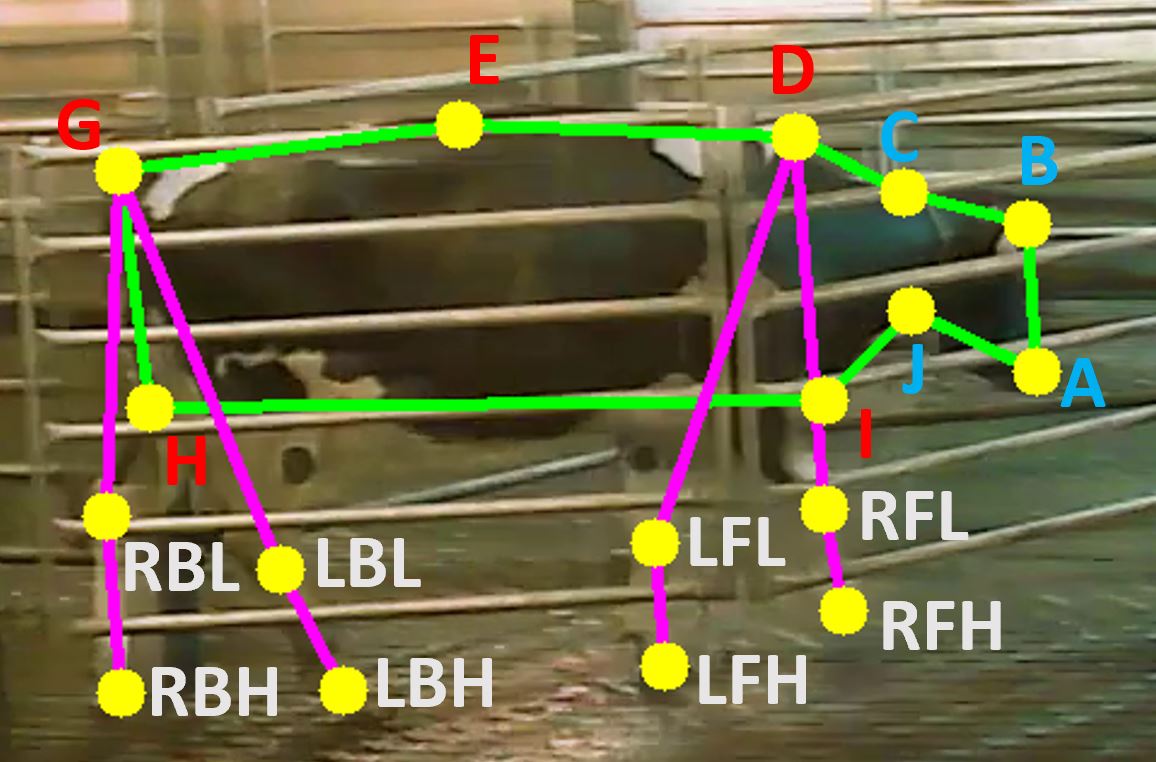}
    \caption{The proposed cow structural model. 4 blue head region points: A:nose, B:head, C:top of neck, J:bottom of neck. 5 red body region points, D:shoulder, E:spine, G:tailhead, H:mid-thigh, I:bottom of shoulder. 8 white leg and hoof points, with name format: Right/Left + Front/Back + Leg/Hoof. }
    \label{fig:cow_skeleton}
\end{figure}

In this paper, we combine deep learning with domain knowledge about cows to develop a cow structure detection system that operates on videos captured from a practical dairy farm.
Our system estimates the number of cow objects in a frame and detects the body parts of every individual cow.
For each cow object, the detected body parts compose a side-view cow structural model as shown in Figure \ref{fig:cow_skeleton}.
This model describes both the spatial location of the cow and additional structural information such as the body contour, the positions of major joints, and the trajectories of their movement.
This detailed information provides interpretable knowledge for further health analysis.

This method also overcomes some practical issues.
First, all the videos are captured without interfering with the daily work on the dairy farm, requiring only surveillance cameras and no specialized hardware.
Second, by incorporating domain knowledge about cows, the video processing algorithm overcomes practical video challenges, such as poor video quality, bad lighting conditions, and heavy occlusion from fences.
Later experiments show that our method provides robust results under these conditions.

There are three main contributions in this work.
First, we design a cow model with keypoints that presents the structural information about a cow that enables further interpretation for subsequent cow health analysis.
Second, a system is developed to extract the cow structural models from videos that are captured from practical dairy farms. 
Third, for the cow structural model, we also develop corresponding evaluation metrics that operate with limited ground-truth labels.
In later experiments, we use multiple video datasets captured from different cameras to test the robustness of the proposed detection system.
This system is also compared with other popular object detection algorithms and demonstrates clear  advantages when presented with practical challenges.
We note that dairy cows are just one type of four-legged livestock, and we anticipate our method can be easily extended to similar animals.

This paper is organized as follows.
Section \ref{sec:previous_work} reviews previous object detection methods and visual-related applications for cows.
Section \ref{sec:model} introduces the proposed cow structural model including the keypoints and their spatial constraints.
Section \ref{sec:method} presents the cow structure detection system, followed by  detail explanations of the detection module and the post-processing module.
The related cow structure evaluation metrics are described in Section \ref{sec:metric}.
Next, Section \ref{sec:experiment} describes three experiments of our detection system:  evaluation of the individual components of our system, performance on different datasets, and a comparison between different methods. 
These three experiments demonstrate the effectiveness of each component of our system, the robustness of our system, and the advantages of our method relative to existing methods, respectively.
Finally, Section \ref{sec:conclusion} summarizes this work.


\section{Previous work}\label{sec:previous_work}

Object detection is one of the most popular topics in image and video processing. 
Traditional Video Object Segmentation (VOS) methods such as \cite{wang2015saliency, tsai2016video, lee2011key} detect objects using motion information from video sequences.
With the development of Convolutional Neural Network (CNN), new learning-based methods achieve much better results.
Methods such as Mask R-CNN \cite{he2017mask}, DeepLab \cite{chen2017deeplab}, and You Only Look Once (YOLO) \cite{redmon2018yolov3}, are widely applied to solve the problem of image semantic segmentation, which requires detection and classification of the objects in an image.
CNNs are also applied for video object segmentation.
One Shot Video Object Segmentation (OSVOS) \cite{caelles2017one} does an online fine-tuning process which is trained on one frame of the video sequence and applied to all the other frames in the sequence.
This method is further extended with image-based detection methods for semantic guidance \cite{maninis2018video}.
Other methods such as \cite{tokmakov2017learning, cheng2017segflow, voigtlaender2017online} apply different models using either temporal information or memory for object detection.
There are also some popular public VOS datasets available for benchmarks, such as the YouTube VOS \cite{xu2018youtube} and DAVIS dataset \cite{perazzi2016benchmark}.
However, all these methods generate bounding boxes or pixel-level masks to represent detected objects, but the structural information of the object is not identified.
Additional processing would be required to extract further detailed information from these masks.

Apart from spatial segmentation, there are also research focusing on object structural information such as keypoint detection and pose estimation.
Benefiting from the public human pose datasets such as MPII \cite{andriluka14cvpr} and COCO human skeleton \cite{lin2014microsoft}, advanced methods are developed for human skeleton detection.
DeepPose \cite{toshev2014deeppose} first applies CNN for human body parts detection based on images, and the stacked hourglass network \cite{newell2016stacked} extends it to detect humans at multiple spatial scales.
To solve multi-human detection, the relationship between human joints are considered. 
ArtTrack \cite{insafutdinov2017arttrack} generates a simplified human body-part model; OpenPose \cite{cao2017realtime, cao2018openpose} and Deepcut \cite{pishchulin2016deepcut} use part affinity fields to model the joint relationships.
However, all these methods are designed by incorporating different levels of knowledge about the human body, and they are not easily altered or fine-tuned for other objects like cows.

Recently, new methods such as the DeepLabCut \cite{mathis2018deeplabcut} toolbox, LEAP \cite{pereira2019fast} and DeepFly3D\cite{gunel2019deepfly3d} extend keypoint detection to animals. 
One advantage of these methods is that they provide a means for users to define body parts; this allows the algorithm to adapt  to different animal structures.
The DeepLabCut toolbox also provides simple access to fine-tune the networks, and it can achieve promising results with a small amount of training data.
However, there are two major limitations of these methods.
First, they only support the labeling of one object per frame, and they do not work with multiple objects. 
This limits their usefulness in many situations.
Second, they are designed for video sequences that have been captured under laboratory conditions, with clean background and clear illuminations.
In later sections, our experiment shows that the DeepLabCut method does not work well on our cow videos.

Apart from general detection algorithms, there are some previous work on cow-related visual applications; most focus on lameness detection.
However, their processing techniques are developed for a specially-designed environment where the captured images are clear enough to process.
Normally their detection targets are limited to a specific region instead of the entire cow body structure.
For example, methods like \cite{poursaberi2010real, viazzi2013analysis} only detect the cow's back curvature while other applications such as \cite{song2008automatic, zhao2018automatic} only track the trajectories of the legs and hooves.
As a result, these methods are not general enough to provide a complete body structure.
Apart from lameness, more researchers focus on cow identification with visual data \cite{ter2017bootstrapping,andrew2017visual}.
Their target is to extract cow features such as traditional image features \cite{zhao2019individual} or CNN-based features \cite{shao2019cattle}, to distinguish different cow identities.
But all these methods need a fundamental step that detects and locates the cows within the images or videos.

\section{Structural cow model}\label{sec:model}

In this section, we first introduce the keypoints in our cow structural model in detail, and then describe the spatial constraints between the keypoints.
These constraints are further used in the detection system for separating multiple cows and detecting missing parts.

\subsection{Cow body keypoints} \label{sec:model_cow}

This proposed structural model is designed to represent a detected cow object in the frame more effectively than using a binary cow mask.
It is designed to provide both the spatial location and cow structural details, such as the body shape and positions of the body parts.
For consecutive video frames, this model should also provide information so that we can track the movement or motions of these body parts.
Inspired by recent approaches to model the human skeleton \cite{toshev2014deeppose}, we combine some anatomical cow joints with other spatial keypoints to represent the cow pose, and the cow structural model is built by connecting the keypoints.

Figure \ref{fig:cow_skeleton} shows our proposed side-view cow structural model.
There are 17 points in total to describe the important locations of a cow object from this angle.
The upper body region has 9 points, including the head region (blue) and the main body region (red).
Connecting these points forms the contour of the upper body region (green lines).
Another 8 points are in leg-hoof regions which represent the four limbs, and each limb has a pair of leg and hoof joints.
Comparing to the anatomical 51-point cow skeleton model \cite{aujay2007harmonic}, we only select visually-observable joints.
Some joints such as the elbow and stifle joints are neglected because their positions are not readily visible and thus difficult to isolate visually.
In addition, we also add some points such as the two bottom corners H and I show in Figure \ref{fig:cow_skeleton}.
Even though they are not physical joints, connecting them with other joints forms a closed contour which spatially locates the body region.
The point E on the spine is also an added point, because connecting three spine points provides information about the back curvature which is useful for lameness detection.

There are two general observations about the keypoints in this cow structural model.
First, the points in the main body region (red in Figure \ref{fig:cow_skeleton}) are always visible from the side-view, and their relative spatial locations do not change dramatically when the cows are walking.
Second, the leg and hoof points are more difficult to detect compared to the upper body region points because of the practical issues such as bad illumination, shadows, and fast leg movement.
Distinguishing between the points from the left or the right leg is also difficult when there are obstacles in front, for example the horizontal fences shown in Figure \ref{fig:cow_skeleton}.

\subsection{Keypoint constraints}

Practical constraints limit the potential relationships among the keypoints in both space and time.
When the cows are walking between the fences, the cameras located at a fixed position on the side wall always capture the side-view of the cow.
In this case, all the cows shown in the video have the similar pose, and the keypoints of their upper body region are always located at relatively fixed spatial positions. 
For example, the cow's head always appears on the right side of the body, and the body does not change size.
As a result, we can compute general relationships that constrain the keypoints in the cow structural model.

To model the constraints, we first define the center of the cow's body.
This center point is computed as the spatial center of all the keypoints from the cow's upper body region.
Note that the points in the leg-hoof region are not used to compute the center point because their positions are not relatively fixed when the cow moves.
Then we can estimate the relative spatial relationship between the center and all the keypoints.

\begin{figure}
    \centering
    \includegraphics[width=6cm]{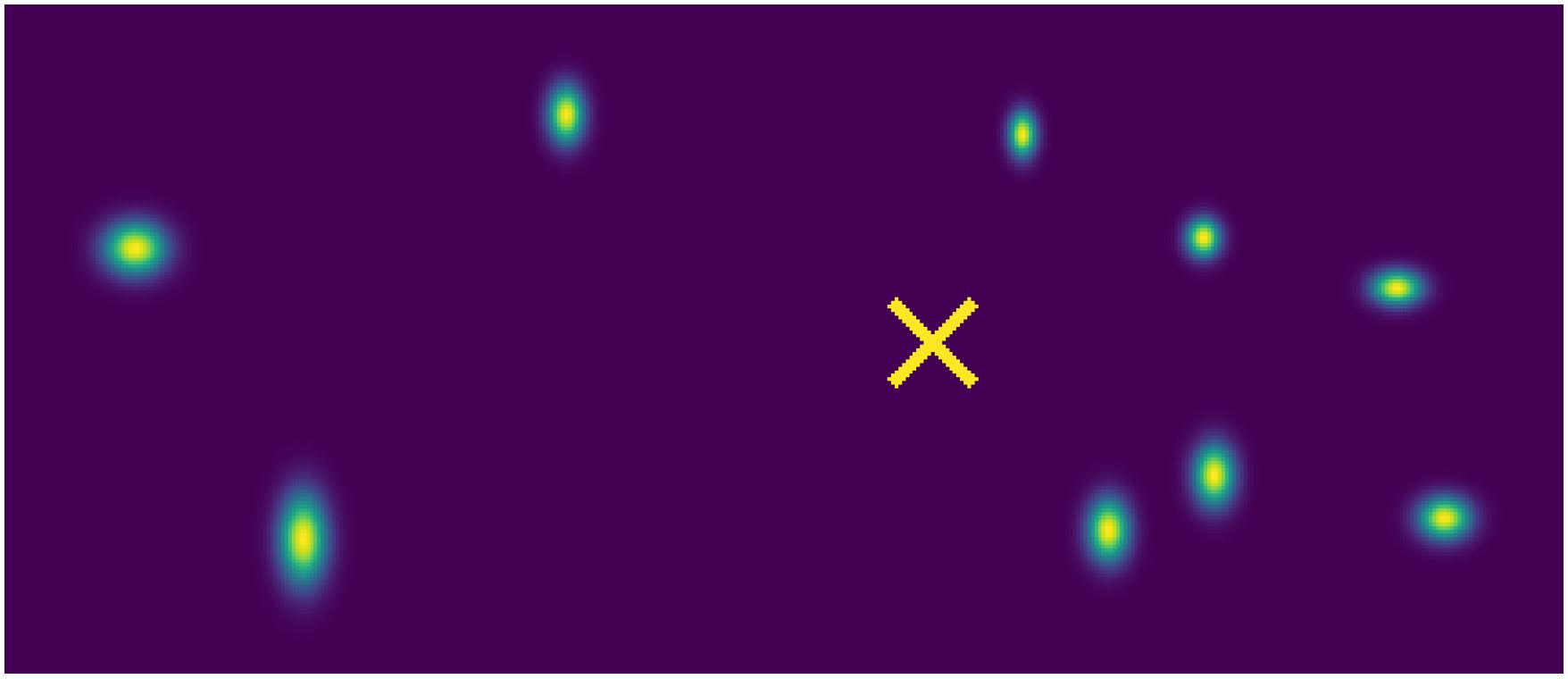}
    \caption{The constraints between the upper body keypoints of the cow structural model. The yellow \textit{X} is the cow center, and the surrounding points shows the relative positions of the keypoints in the upper body region.}
    \label{fig:cow_body_model}
\end{figure}

Figure \ref{fig:cow_body_model} visualizes the keypoint constraints.
The middle \textit{X} shows the cow center $c$, and the relative spatial locations of the upper body parts appear surrounding the center. 
Notice each body part mapping function $F_{j}$ is a 2D Gaussian probability distribution, which is shown as the ellipse in the figure.

Formally, for a fixed cow center point $c^{*} = (x, y)$, we define a set of mapping functions $F_{j}(\cdot)$ that describe the relative spatial locations of every upper body-part point $p^{*}_{j}$ to the center,
\begin{equation}\label{eq:body_center}
    p^{*}_{j} = F_{j}(c^{*})
\end{equation}
where $j$ is the index of the body part.
Each mapping function $F_{j}$ is characterized by a 2D Gaussian model, and the parameters are trained using all ground-truth labels.
During the training process, the approximate cow center $c$ is computed first by averaging all labelled body parts, and the parameters in each $F_{j}$ are estimated individually based on their relative spatial locations to $c$.

In the next section, we show how these constraints can be used to separate cows which are spatially close together in the frame. 
They also provide a reference when assigning body-part candidates to each individual cow object in the post-processing module.

\section{Skeleton detection system}\label{sec:method}

This section introduces our proposed system to detect the structure of cows.
We first review one popular work for keypoint extraction and then describe the components of our proposed system. 
Then we explicitly introduce two main processing components: the body part extraction module and the post-processing module.

\subsection{The DeepLabCut toolbox}

The DeepLabCut toolbox \cite{mathis2018deeplabcut} is a recent popular method to extract keypoints from video sequences.
The inputs are color images from videos, and it applies a CNN to generate confidence maps that represent the potential keypoint locations.
One advantage of the DeepLabCut toolbox is that it provides simple access for users to manually define the output body parts, and the toolbox automatically alters the last layer of the CNN based on the number of body parts.
For example, there are 17 confidence maps generated in our case because we have 17 keypoints in our cow structural model.
In our system, we apply the network created by the toolbox to extract the keypoints of our cow structural model.

However, other modules from the toolbox are less suitable for our application because of two major limitations.
First, this platform is designed and evaluated with videos captured from a laboratory environment with clear objects and background.
But our cow videos, generated from a commercial farm, have low video quality and the view of the cows are often blocked by obstructions.
Later experiments show that the original DeepLabCut does not provide robust detection results on our videos.
Second, this method assumes there is only one object in a frame, so it only chooses one body part from each confidence map.
If there are multiple body-part candidates detected, only the position with the highest confidence score will be selected.
But in videos generated from commercial farms, there could be multiple cows and obstructions like fences that easily cause false detection.
We address these two limitations and build a general keypoints detection system which extracts robust keypoints on our cow videos.

\subsection{Proposed system}

\begin{figure}
    \centering
    \includegraphics[width=8.8cm]{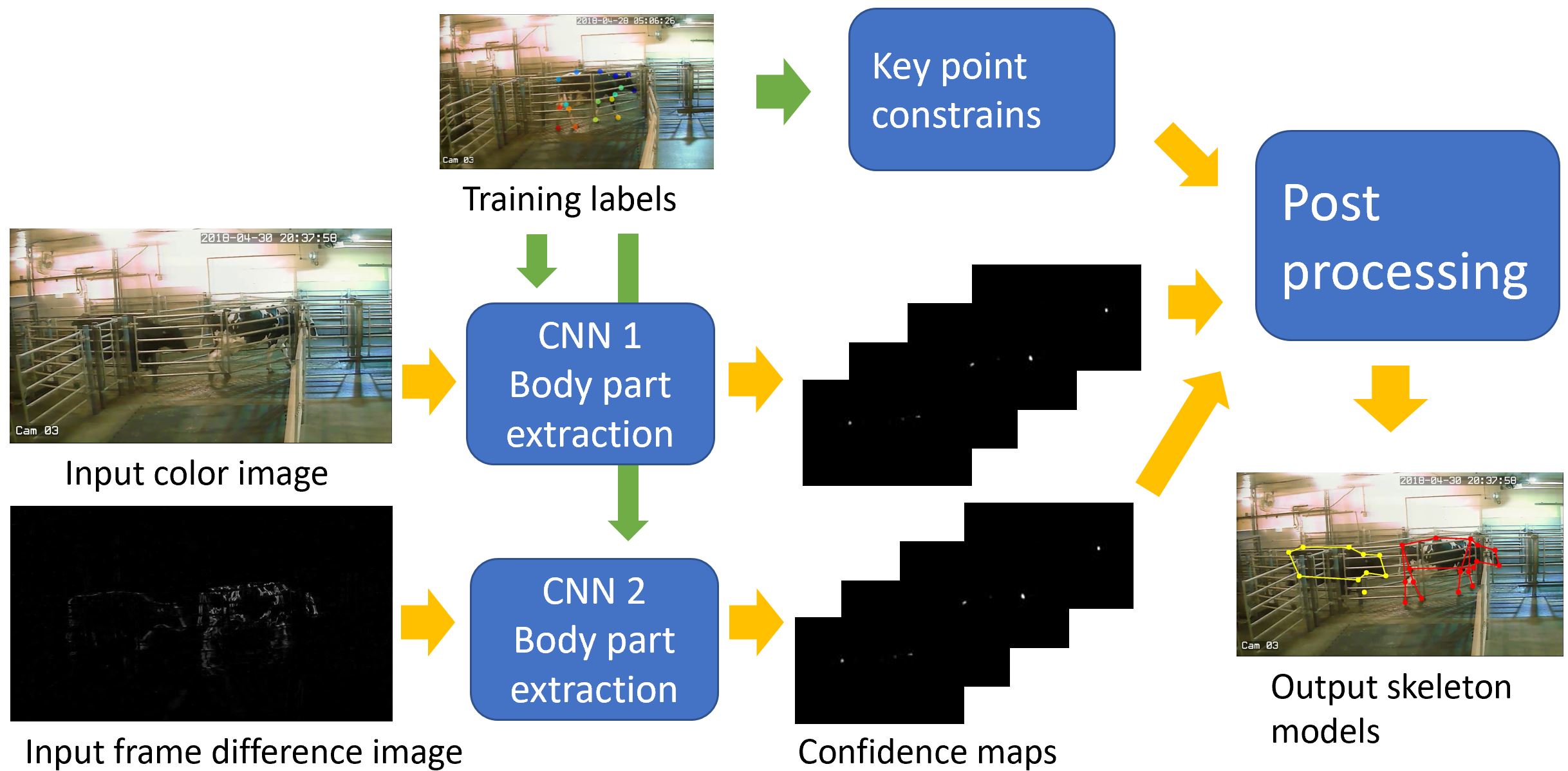}
    \caption{A diagram of the proposed system. The green arrows show the training process and yellow arrows present the process during operation.}
    \label{fig:flowchart}
\end{figure}

This detection system is targeted to extract the structural model for every cow object from video sequences.
Figure \ref{fig:flowchart} presents the overall system; its primary components are two CNNs for the extraction and a post-processing module.
The body part extraction module uses trained networks to convert each single image into a group of confidence maps.
Each map shows the potential locations of a particular body part, and the values of the map represent their detection confidence.
The post-processing module generates the final structural model based on two groups of confidence maps and the trained keypoint constraints.
Both modules are discussed in detail in the next two sections.

In this figure, both the training process and the testing process are labelled using colored arrows.
During the training process (indicated by the green arrow in Figure \ref{fig:flowchart}), the ground-truth labels are used to fine-tune both CNNs and the keypoint constraints.
During operation (indicated by the yellow arrow), the system takes the input of both the color image and the frame difference image on the left, and generates the cow structural model for a single image.
After all the frames from a video sequence are processed, the post-processing module refines all the detected cow structures based on temporal information.

\subsection{The body part detection module}\label{sec:method_dlc}

The goal of this module is to find the spatial locations of all potential keypoints from raw images.
In our system, we apply the original DeepLabCut network \cite{mathis2018deeplabcut}, labeled CNN1, to extract keypoints from color images. 
This network structure follows DeeperCut \cite{insafutdinov2016deepercut}, and is implemented using ResNet \cite{he2016deep} for the convolution stages, followed by one de-convolution layer before the output layer to recover the target spatial locations of the keypoints. 
The last two convolution layers apply atrous convolution, which increases effective fields-of-view of the applied convolution and preserves spatial resolution \cite{chen2017deeplab}.
By default, the DeepLabCut network is pre-trained on ImageNet \cite{krizhevsky2012imagenet} for image classification tasks, and we use our own cow labels to fine-tune the last de-convolution layer for keypoint detection.

However, as mentioned above, low video quality and heavy obstacles influence the performance. 
To overcome this issue, we add an extra network, CNN2, into the system.
The architecture of this network is same as the first, but it processes frame difference images.
There are three major advantages of using frame difference images for our cow videos.
First, because we have fixed cameras, the frame difference image better captures the moving objects and eliminates the stationary obstacles such as fences.
Second, many of our target keypoints are on the contour of the cow body, and the frame difference highlights these edges of a walking cow.

Third, frame difference also reduces the influence of color variation.
This is useful, because the color responses of different cameras are not the same especially under poor illumination. 
In addition, the majority of the cows have color variations introduced by the patterns on the cows, but some cows only have a single coloring, such as pure white, black or brown..
If these patterns are not included in the training frames, then the color-based CNN methods would likely fail to detect cows with unseen colors.
As a result, using frame differences provides robustness to these factors.

However, using the frame difference images alone is not enough because they eliminate too much spatial information, especially for legs and hooves.
This is because most of the legs are stationary even when the cows are moving.
As a result, our system merges both networks together to improve the body part detection accuracy.

\subsection{The post-processing module}\label{sec:method_pp}

The post-processing module collects and merges the confidence maps from the two CNNs, and assigns the cow body-part candidates to each cow object instead of just to one cow per frame.
This step enables the system to detect multiple cows together and track their temporal movements.
There are three major steps in this post-processing module: body part extraction, spatial clustering, and temporal filtering.

\subsubsection{Body part extraction}

This step extracts the spatial locations of all body-part candidates from the confidence maps generated by the CNNs. 
Notice that at this stage, the number of cow objects in the image is unknown and we want to extract all possible candidates.
For each body part, the confidence map from the two networks are merged together, and we use non-maximum suppression to select all the points whose confidence scores are higher than their neighbors.

The output of this step are lists of body-part candidates.
Formally, for a given frame at time $t$, all these body-part candidates can be represented as $p_{j}^{i, t} = (x, y)$, where $j$ is the index of that body part, and $i \in \left \{1,2 ...  \right \}$ indicates the count of all possible keypoints extracted for this body part. 
The total number of $i$ is not determined because the number of cow objects is unknown at this stage, and there could be some incorrectly-detected candidates.
All these candidates are further selected and clustered in the next step.

The confidence maps from the two networks are treated differently during this process.
For keypoints from the upper body region, the two confidence maps are directly merged to find body parts. 
But since color information is more useful to detect the legs and hooves, only the confidence map from the color image network is used unless this map contains no candidates.

\subsubsection{Spatial clustering}

The second step in the post-processing module is spatial clustering.
This step selects the correct body parts and clusters them into different cow objects.
The first task before clustering is to determine the number of cows in the frame by counting cow centers.
Given a set of extracted keypoint candidates $p_{j}^{i, t}$ from the upper body parts, the corresponding cow center positions can be estimated based on the constraints of the keypoints, shown in Equation \ref{eq:inverse_body_center}.
\begin{equation} \label{eq:inverse_body_center}
     c_{j}^{i, t} = F^{-1}_{j} (p_{j}^{i, t})
\end{equation}
Then a mean-shift clustering method is applied to the 2D spatial positions of all the cow centers $c_{j}^{i, t}$.
Based on the clustering results of the center points, the corresponding body parts are labelled into separate cow objects.
We ignore a cow object if the system cannot detect more than half of its keypoints.

The cow centers are also used to predict the location of missing body parts that the network fails to detect.
After all keypoints are clustered into distinct cow objects, then for each cow object, we compute the averaged cow centers based on the detected points, and the miss-detected keypoints can be estimated using the keypoint constraints $F_{j}$.
The predicted body parts based on the constraints may not be always accurate, but they provide a rough estimate of the cow's spatial location, which is useful for searching for keypoints in leg-hoof regions.

The final process in this step is to match the leg-hoof points.
Similar to \cite{zhao2018automatic}, we indicate the region of all possible leg-hoof points using a rectangle that is one-third wider than the rectangle of the upper body.
Candidates outside this region will not be considered.
The search process relies on the structural model.
We follow the order of \textit{shoulder/tailhead}, \textit{leg}, \textit{hoof} along each limb, and search the joints from among the candidates that lie in the search range.
We also reject inappropriate points by applying the rule that each limb should have a certain rotation range; the angle between \textit{shoulder} to \textit{leg} and \textit{leg} to \textit{hoof} must be greater than 90 degrees for valid keypoints.
Finally, all the selected leg-hoof joints are connected to the body contour to complete the final cow structural model.

\begin{figure}
    \begin{subfigure}[b]{.49\linewidth}
        \includegraphics[width=\linewidth]{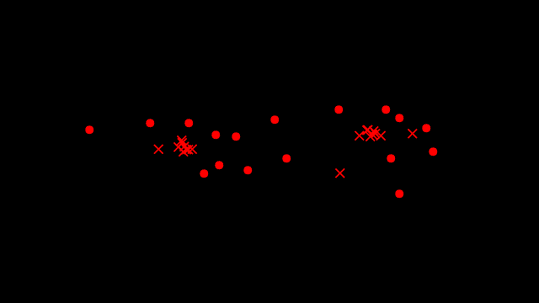}
        \caption{The keypoints and centers}
        \label{fig:pp_process_p1}
    \end{subfigure}
    \begin{subfigure}[b]{.49\linewidth}
        \includegraphics[width=\linewidth]{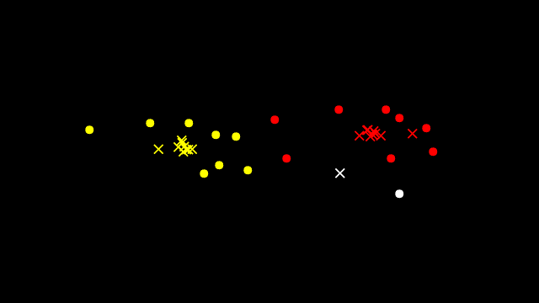}
        \caption{Center points clustering}
        \label{fig:pp_process_p2}
    \end{subfigure}

    \begin{subfigure}[b]{.49\linewidth}
        \includegraphics[width=\linewidth]{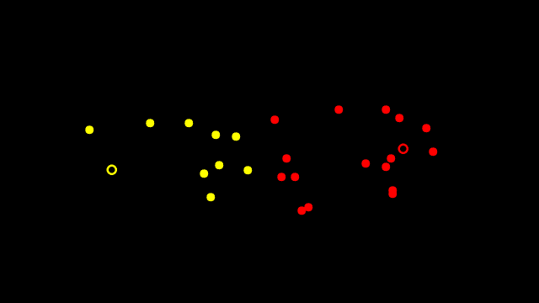}
        \caption{Adding leg region points}
        \label{fig:pp_process_p3}
    \end{subfigure}
    \begin{subfigure}[b]{.49\linewidth}
        \includegraphics[width=\linewidth]{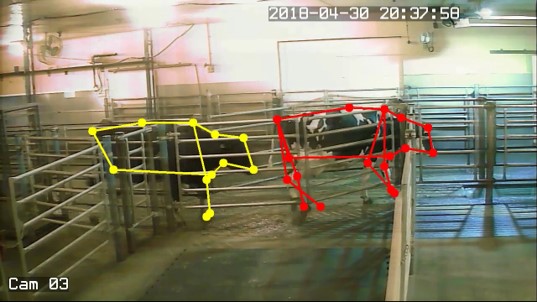}
        \caption{The output structural model}
        \label{fig:pp_process_p4}
    \end{subfigure}
    \caption{The procedure of spatial clustering during post-processing. Circles represent the body parts $p$ and crosses are the estimated cow centers $c$. Empty circles are the predicted body parts. Each color indicates a different cow object.}
    \label{fig:pp_process}
\end{figure}

Figure \ref{fig:pp_process} illustrates the procedures of the spatial clustering step.
The top left image shows the original extracted body parts from the previous step.
The red circles are the extracted candidates and each is converted to a corresponding cow center, shown as crosses.
Then in the top right image, all center crosses are clustered using mean shift to produce three clusters shown in distinct colors.
Here the incorrect cluster (white) is eliminated because there are not enough candidates.
Next in the bottom left image, points in the leg and hoof region are assigned to each cow object.
Notice the empty circles are predicted points; the yellow one is blocked by the fences.
Finally, by connecting all keypoints together, we form two cow structural models as shown in the bottom right image.

\subsubsection{Temporal filtering}
The final step in post-processing module refines the detection results using temporal information and matches cow objects across different frames.
The two previous steps each operate on a single image, but the relationship between neighboring video frames is helpful to refine keypoint positions. 
It is reasonable to assume that the cows walk on an identical path between the fences and that they move steadily and slowly.
This means that for a specific keypoint in the upper body, its trajectory over time should be smooth and any points far from the trajectory line can be considered outliers.

Based on this idea, we refine the positions of every upper body-part point across time to improve the temporal smoothness of the output.
Before this step starts, all the frames in a video have been processed, so we know the number of cow objects in each frame.
Then for every body part in the upper body region, we temporally filter each trajectory to remove and correct the outliers.
In our experiment, we use a median filter, which is simple and provides robust prediction.
Other filters such as the Kalman filter do not work well especially when there are too many missing points from the previous steps.
Notice that the leg-hoof region points are not involved in this process, since their trajectories are much more complicated.

Based on the trajectories of each cow object, the cow objects can be matched between neighboring frames. 
After this process, the system detects the total number of cows shown in a complete video sequence, and parameters about how every cow moves can be inferred, including the speed and rhythm \cite{whay2002locomotion}.

\section{Evaluation metrics} \label{sec:metric}

This section introduces our evaluation metrics.
Although our method uses few ground-truth labels for training, ground truth is typically also required for performance evaluation.
Therefore, in this paper we propose to use both supervised measures, which compare the detected results with ground-truth labels, and unsupervised measures, which directly evaluate the results without labels.
Adding unsupervised measures to the evaluation process improves its thoroughness in the presence of insufficient labels.
We first discuss the supervised measures for the cow structural model, and then introduce two unsupervised metrics.

\subsection{Supervised measures}

Quantifying the performance of the cow structural model requires more than the typical measures used to quantify object detection.
As mentioned in Section \ref{sec:intro}, the cow structural model is designed to provide two types of information: the spatial location of the body region, and the detailed positions of body parts.
Both information is represented in terms of the keypoints of the cow body parts, and our ground-truth labels are also in terms of keypoints.
As a result, we separately evaluate the area of the cow body region and the points in the leg-hoof region.
Two metrics are developed and described below in detail: the \textit{Body F1 score} and the \textit{Leg-hoof F1 score}.
In each case, the F1 score is harmonic mean of precision and recall when comparing the detection results to the ground truth. 
Notice that both metrics compare accuracy at the keypoint level.
In a later experiment in Section \ref{sec:experiment_mask}, we also propose a method to convert the cow structural model to a binary mask with both body region and extended limbs, for the sole purpose of comparing our detected keypoint model with other mask-based segmentation methods.

\subsubsection{Body F1}
This metric measures the spatial area formed by the body region points.
We connect the keypoints in the upper body region and generate one polygon mask for both the detected structural models and the ground-truth keypoints.
Then we compare the two masks using the typical Intersection Over Union (IOU) metric and report the F1 score.

\subsubsection{Leg-hoof F1}

For the legs and hooves, a single pixel position represents each keypoint. However, physical joints typically extend for a larger spatial region. Therefore, the evaluation metric must accommodate this discrepancy, which may introduce systematic errors to both the labelling and detection process.
For this reason, when measuring the distance between ground truth and the detected leg and hoof keypoints, we set a threshold distance of 30 pixels.
If the distance between the points is less than this threshold, we consider the joint to be detected, and points further away are considered to be miss-detected.
After thresholding, we determine how many leg-hoof points are successfully detected, and summarize this using the F1 score computed from the precision and recall.
Since we do not create ground-truth labels for keypoints that are completed blocked by obstacles, these blocked joints do not affect the evaluation result.

\subsection{Unsupervised measures}

Unsupervised measures allow performance evaluation without ground-truth labels. 
This is particularly critical for video, where exhaustive application-specific labeling becomes even more onerous. 
Without labels, previously proposed metrics such as mean of region similarity, contour accuracy \cite{li2017video}, and temporal stability metric \cite{perazzi2016benchmark} cannot be computed.
Here, we apply prior knowledge to evaluate the performance when the ground-truth labels are not provided.


We consider two rules for the cow structural model.
First, the spatial locations of the keypoints in a model should always form a cow-shaped object.
Second, the shape of the cow body should be stable during the walk and the keypoints should have similar smooth trajectories.
Based on these two constraints, we introduce two unsupervised metrics: the valid cow percentage and temporal consistency.

\subsubsection{The Valid Cow Percentage (VCP)}

This metric counts the fraction of detected cow models that are valid.
Here valid means that the positions of the keypoints in the structural model can form a cow-shaped object.
Like the supervised measure, we validate the upper body region and leg-hoof region separately.

For the upper body region, we use the trained keypoint constraints (Figure \ref{fig:cow_body_model}) as a reference, and compute the similarity between the detected contour and the reference using the Fr\'echet distance \cite{alt1995computing}.
We choose this distance because it better captures the similarity between two curves, which are the body contour in our case.
The computed distance is thresholded to form a binary decision whether the upper body region is valid or not.
For points in the leg-hoof region, we define two interpretable rules to validate their spatial positions: all leg-hoof points should be lower than the body region points, and all hoof points should be lower than their corresponding leg points.
If all leg-hoof points satisfy these two rules and the upper body region contour is also validated, the cow structure is considered valid.

This validation scheme is applied to all the detected cow objects in a video sequence, and the Valid Cow Percentage (VCP) is computed as the number of valid cow objects divided by the number of detected cows. 
The absolute VCP score is directly related to the actual number of cows in the testing video sequence, so the score is only meaningful when compared with other methods on the same testing dataset.

\subsubsection{Temporal Consistency (TC)}

The second unsupervised metric evaluates the Temporal Consistency (TC), which reflects the smoothness of the motion of moving objects in a video sequence.
It is reasonable to assume that at a certain camera angle, the points from the body region always share the same translational motion because the shape of the cow body is stable.
So ideally, the motion vector between every keypoint generated from one frame to the next frame should be the same.
The Temporal Consistency (TC) metric evaluates this co-movement and computes the difference between the motion vectors generated by the body parts.

Formally, for each body part $p_{j}^{t}$ in a cow object, we compute its motion vector from time $t$ to $t+1$ and summarize the variations $d$ between all the motion vectors as
\begin{equation}
    d^{t} = std(p_{j}^{t+1} - p_{j}^{t}), \forall j \in \left \{j_{1}, j_{2} ...\right\}
\end{equation}
where $std$ is the standard deviation, and $j_{i}$ represents the index of the body part from the upper body region.
Then the temporal consistency is computed as the average motion vector differences for all the frames in a video sequence. 
\begin{equation}
    TC = mean(d^{t}),  \forall t
\end{equation}
Notice this measure is applied to every individual cow object in a video sequence, and smaller TC values imply smoother object movements.

\section{Experiments}\label{sec:experiment}

This section presents the validation experiments.
We first give a high-level summary of how we collect and prepare the video data from a commercial farm.
Then we present three different experiments.
The system output experiment compares the results of every stage in the proposed method, to demonstrate the importance of each component.
Next, the dataset robustness experiment is performed on three different sets of video, to demonstrate the robustness of the method.
Finally, we compare our method with other popular object segmentation methods, to demonstrate the advantages of our proposed method for cow detection.


\subsection{Data collection}\label{sec:experiment_a}

All cow videos in these experiments are collected from the Purdue Animal Sciences Research and Education Center located in West Lafayette, IN, USA from 2018 to 2019. 
All procedures were approved by the Institutional Animal Care and Use Committee (PACUC \#1803001704).
The cameras are mounted at fixed positions include a side-view of the path where cows walk every day.
This path has fences on both sides and only allows one cow to walk through at a time.
This limits the amount of cow-overlap; however the dense fences partly block the view of the cows, and some body parts are not visible behind the fences, as shown in Figure \ref{fig:cow_skeleton}.
This walking path is a typical component of many dairy farms.

During the course of data collection, we used three different capture devices: a commercial surveillance camera with Digital Video Recorder (DVR), a GoPro camera, and a high-quality IP camera.
Table \ref{tab:data} shows the detailed information of the three video sets captured from the three cameras.
The DVR videos have the worst quality with low frame rate and low resolution.
The GoPro videos provide higher frame rate, but they are spatially cropped with less spatial details.
The IP camera captures high quality videos with both high frame rate and rich spatial information.

Table \ref{tab:data} compares several factors among the cameras that will influence detection performance.
As noted, the video resolution and frame rate are different between the three sets, and Set 3 has the best quality.
The number of pixels per cow refers to the average number of pixels that each cow occupies in an image, which is an indication of the spatial detail in each set.
Notice that Set 2 only has 0.29 million pixels per cow, which is less than a third of the other two sets.
The field-of-view each camera are also different.
Set 1 videos only capture the center of the walking path where there are fewer fences, while the other two sets capture a wider view which includes two sides that have denser fences. 
In addition, the typical number of cows in one video are different across the sets. 
Narrow field-of-view videos normally captures a single cow in the frame, but the wider-angle videos could contain multiple cows, which challenges the detection method.
In general, Set 3 has more video clips than the others with the greatest variety, so we will further divide this set into subsets in a later experiment described in Section \ref{sec:experiment_mask}.

To prepare the videos, we temporally segment the hours-long sequences into 10-second clips, on average, where all cows walk from left to right. 
In each set, we separate the clips into training and testing groups, where the number of training clips per set are shown in parentheses in Table \ref{tab:data} after the number of video clips. 
All multiple-cow clips are testing clips, so the training clips all contain only a single cow object. 
Non-consecutive frames are chosen randomly for labeling from both training and testing clips.

\begin{table}[]
\begin{center}
    \caption{Summary of three sets of video data used in the experiments. The \# Pixel per cow is in units of millions.}
    \begin{tabular}{|c|c|c|c|}
    \hline
     & Set 1 & Set 2 & Set 3  \\ \hline
    Capture Device & DVR & GoPro & IP camera \\ \hline
    Video info & \begin{tabular}[c]{@{}c@{}}1280*720\\ @12fps\end{tabular} & \begin{tabular}[c]{@{}c@{}}1232*384\\ @30fps\end{tabular} & \begin{tabular}[c]{@{}c@{}}1920*1088\\ @30fps\end{tabular} \\ \hline
    \# Pixels per cow & 0.88m & 0.29m & 1.35m \\ \hline
    Image Quality & low & low & high \\ \hline
    Field of view & narrow & wide & wide \\ \hline
    \# cow per clips & single & multiple & multiple \\ \hline \hline
    \begin{tabular}[c]{@{}c@{}}\# video clips\\ (\# for training)\end{tabular} & 87 (5) & 18 (2) & 114 (5) \\ \hline
    \# training frames & 100 & 40 & 100 \\ \hline
    \# testing frames & 585 & 59 & 611 \\ \hline
    \end{tabular}

\label{tab:data}
\end{center}
\end{table}

\subsection{System component evaluation}

This experiment compares all the internal outputs from our proposed system shown in Figure \ref{fig:flowchart}, to demonstrate the importance of each individual module.
We choose the output of CNN1 as the baseline method, which is the original method in the DeepLabCut (DLC) toolbox \cite{mathis2018deeplabcut}.
However, this method can only detect one object per frame, so for a fair comparison, we only use the videos in Set 1 since these only contain one cow object.
We compare the baseline method with four other internal outputs from the system: the CNN2 output from the difference videos, the CNN1 output plus the Post-Processing (PP) stages, the CNN2 output plus the PP stages, and the final merged result.

The implementation details are explained below.
The frame difference images are generated by the sum of differences between the current frame and both the previous and next frame.
The training labels from Set 1 are used to fine-tune both CNNs in the system.
Recall that CNN1 processes the color images and CNN2 processes the frame-difference images.
Both networks are pre-trained on ImageNet \cite{krizhevsky2012imagenet} and their final upsampling layers are fine-tuned with our cow images.
For the two CNN methods without PP stage, we follow the extraction method from the DeepLabCut toolbox by setting a hard threshold and finding the location in the confidence maps with the maximum probability.

Both supervised and unsupervised evaluation metrics are used, but their testing data are different.
For unsupervised measures, we compare the Valid Cow Percentage (VCP) and Temporal Consistency (TC) for all the frames in the testing videos because no labels are required.
But for supervised measures, only the 585 labelled testing frames are used for evaluation. 
Among these labelled images, we report the body F1 score and leg-hoof F1 score, and the VCP score is also computed to compare the cow detection capability of each module in the system.
Both qualitative and quantitative results are presented below.

\begin{figure}
    \begin{subfigure}[b]{.32\linewidth}
        \includegraphics[width=\linewidth]{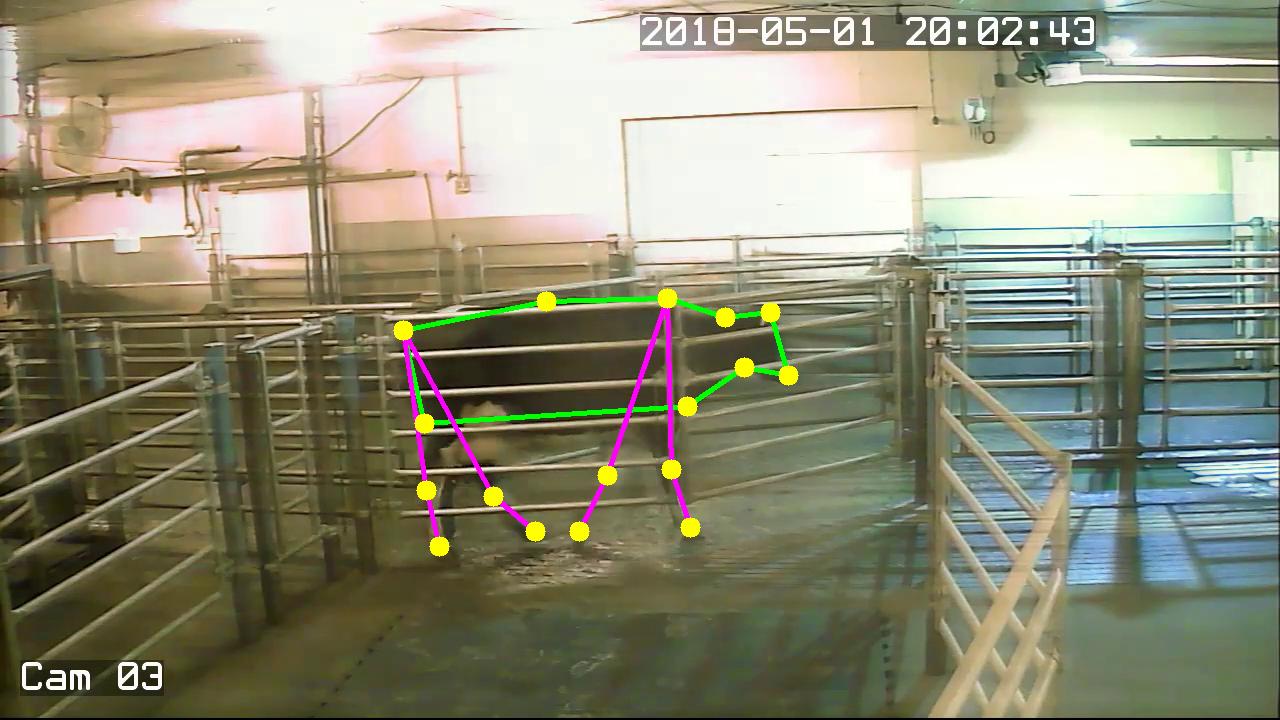}
        \caption{Ground truth}
        \label{fig:exp1_gt}
    \end{subfigure}
    \begin{subfigure}[b]{.32\linewidth}
        \includegraphics[width=\linewidth]{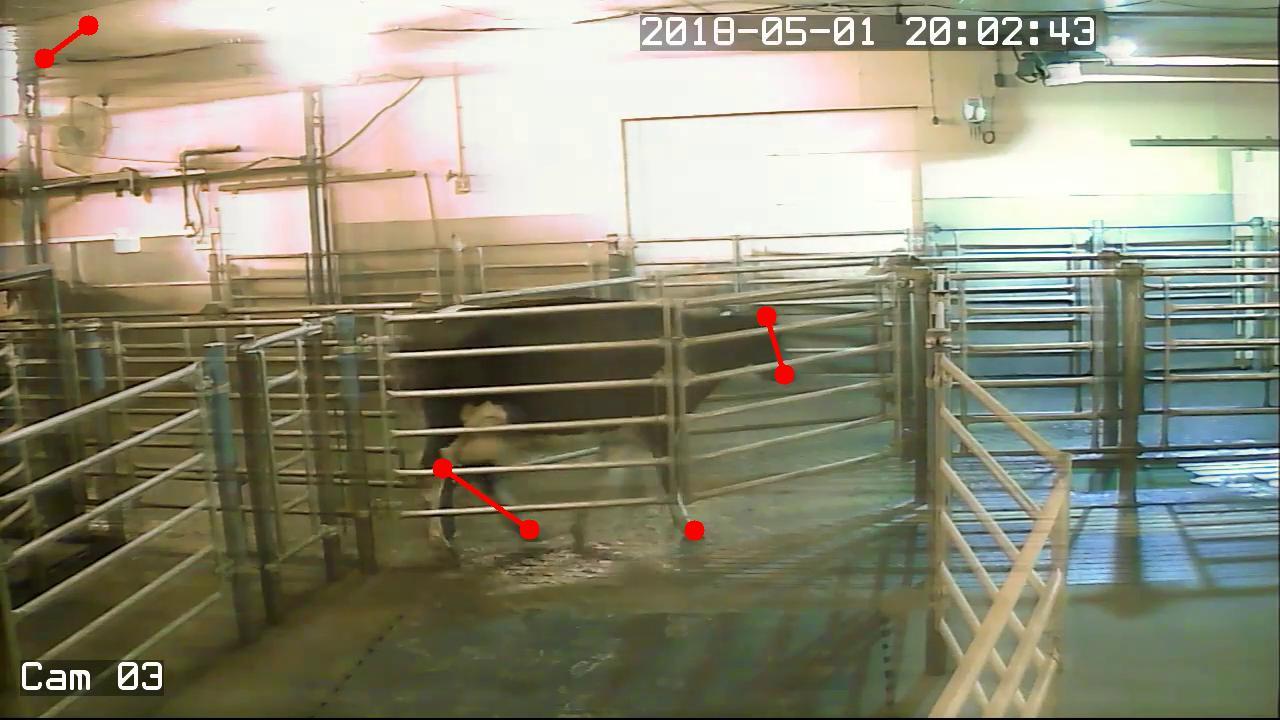}
        \caption{CNN1\cite{mathis2018deeplabcut}}
        \label{fig:exp1_dlc_color}
    \end{subfigure}
    \begin{subfigure}[b]{.32\linewidth}
        \includegraphics[width=\linewidth]{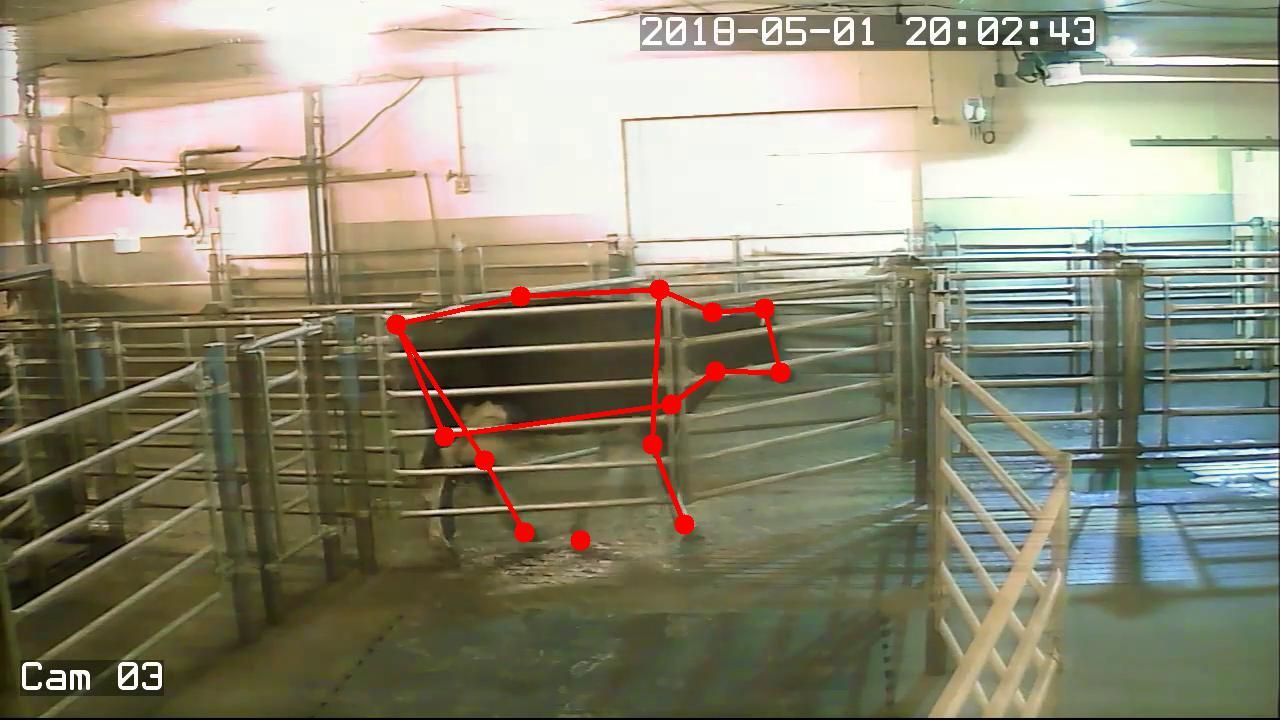}
        \caption{CNN1+PP}
        \label{fig:exp1_dlc_color_pp}
    \end{subfigure}
    \begin{subfigure}[b]{.32\linewidth}
        \includegraphics[width=\linewidth]{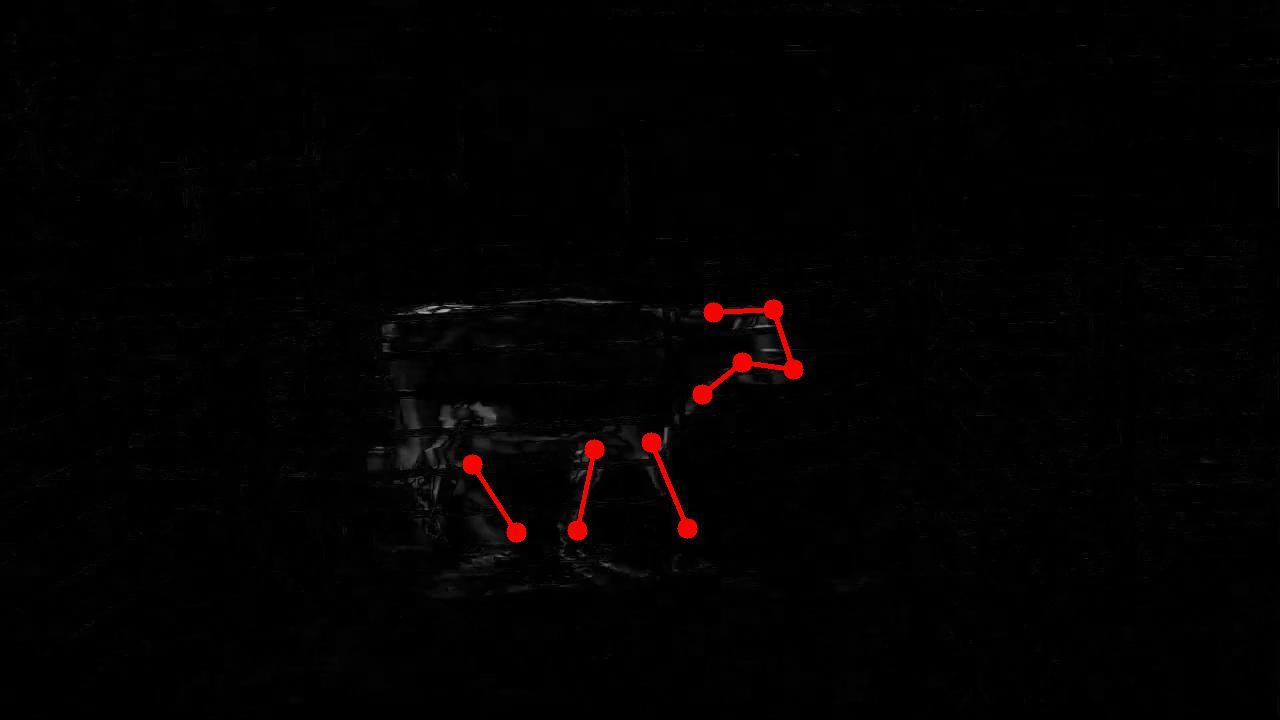}
        \caption{CNN2}
        \label{fig:exp1_dlc_dif}
    \end{subfigure}
    \begin{subfigure}[b]{.32\linewidth}
        \includegraphics[width=\linewidth]{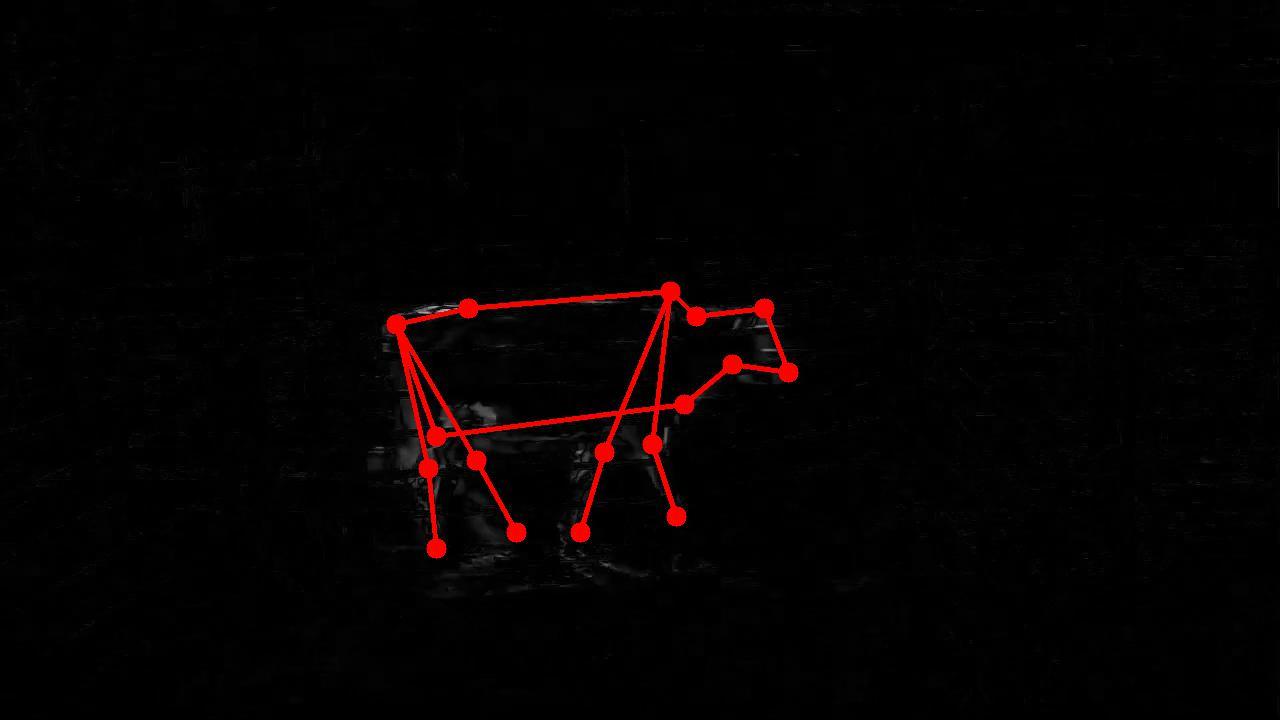}
        \caption{CNN2+PP}
        \label{fig:exp1_dlc_dif_pp}
    \end{subfigure}
    \begin{subfigure}[b]{.32\linewidth}
        \includegraphics[width=\linewidth]{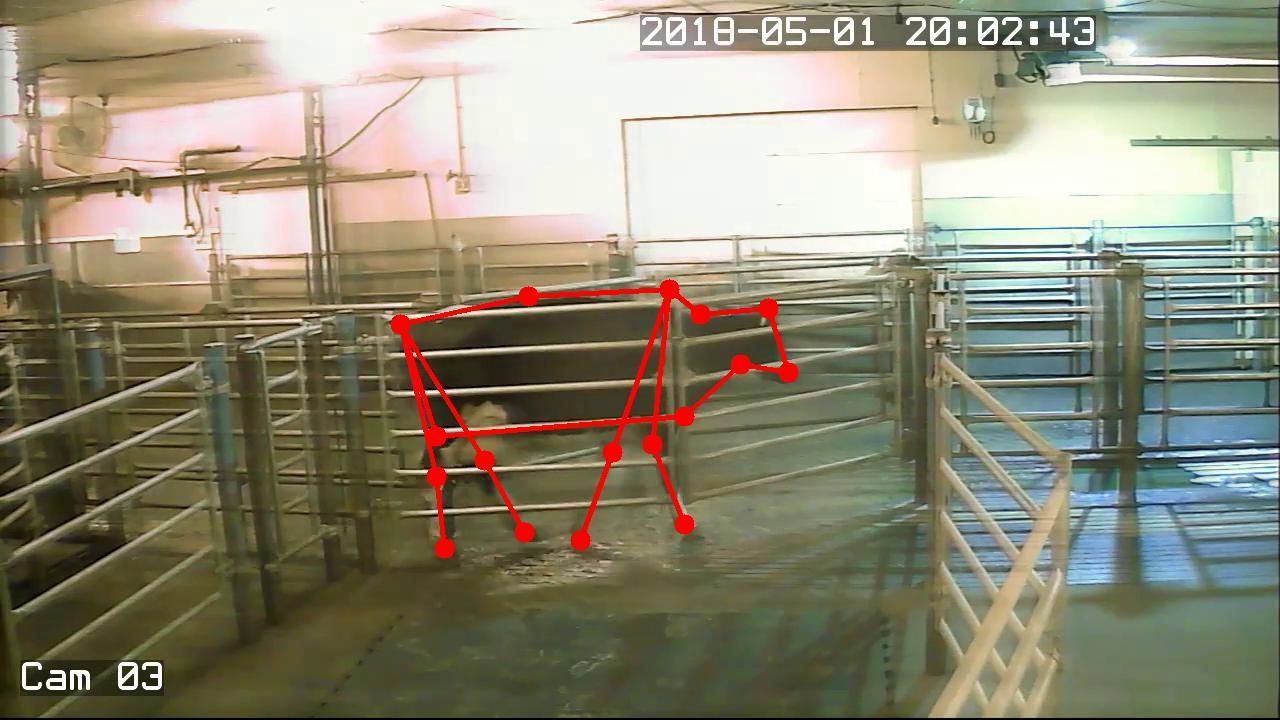}
        \caption{Merged output}
        \label{fig:exp1_merge}
    \end{subfigure}
    \caption{The outputs of different stages in the proposed system.}
    \label{fig:exp1}
\end{figure}

Figure \ref{fig:exp1} shows an example of all five outputs of one testing image in Set 1. 
The direct outputs from the two CNNs without post-processing (top middle and bottom left) miss-detect some body parts, because they apply the strategy from the original DLC method that only selects one maximum point.
Our proposed post-processing module uses non-maximum suppression to select all local maximum values from the confidence map, and all body-part candidates are detected (see bottom right of Figure \ref{fig:exp1}).
Considering the leg-hoof points, some joints of the swing leg are missed by CNN1 based on color image, because of motion blurriness and heavy compression.
But these points are detected by CNN2 using the frame difference image, and the merged result generates a complete cow structural model.

\begin{table}[]
    \centering
    \caption{Comparison of the outputs of the system components on Set 1 videos (single-cow). Notice smaller TC value means smoother object movement in the video. Bold numbers show the best performance method in each column.}
    \begin{tabular}{|c|c|c|c|c|c|}
    \hline
     & \multicolumn{2}{c|}{Unsupervised} & \multicolumn{3}{c|}{Supervised} \\ \hline
     & VCP & TC & VCP & Body F1 & Leg-hoof F1 \\ \hline
    \begin{tabular}[c]{@{}c@{}}CNN1 \\ (DLC{}\cite{mathis2018deeplabcut})\end{tabular} & 0.447 & 102.8 & 0.714 & 0.260 & 0.391 \\ \hline
    CNN2 & 0.408 & 155.0 & 0.673 & 0.366 & 0.252 \\ \hline
    CNN1+PP & 0.632 & \textbf{8.92} & 0.846 & 0.772 & 0.373 \\ \hline
    CNN2+PP & 0.667 & 10.19 & 0.929 & 0.841 & 0.333 \\ \hline
    Merged output & \textbf{0.705} & 9.0 & \textbf{0.960} & \textbf{0.879} & \textbf{0.434} \\ \hline
    \end{tabular}
    \label{tab:exp1}
\end{table}

The numerical comparison results are presented in Table \ref{tab:exp1}.
In general, our complete system (last row) improves the performance compared to the method in the DLC toolbox (first row).
It can be observed that adding a Post-Processing (PP) module largely improves the system performance.
The temporal and spatial prediction in the PP module improves the cow-detection ability demonstrated by the increasing VCP scores.
Notice the two VCP scores from supervised measure and unsupervised measures are not comparable because their test sets are different.
In addition, the temporal filtering process in the PP module largely improves the Temporal Consistency (TC), because the original CNN method purely operates on an image without considering temporal information.
Comparing two F1 scores in the supervised measures, the PP step improves the detection accuracy for the cow structural model because more body-part candidates are selected from the intermediate CNN output.

Comparing the first two rows from the table, we can see CNN2 has better performance than CNN1 for the cow body region but works poorly on the leg and hoof regions.
As explained in Section \ref{sec:method_dlc}, CNN2 operates on gray-scale edges generated by the frame difference and better captures smoothly moving objects like the body region.
But it cannot work in isolation because it eliminates too much information contained in the original images, such as the stationary legs.
As a result, merging the two networks together obtains better detection for the leg-hoof region points.

\subsection{Dataset robustness evaluation}

This experiment evaluates the system robustness with different datasets. 
Training-based detection methods normally perform worse when they are applied to testing data that is substantially different from the training set.
In this experiment, we evaluate the performance of our system when testing on frames collected from the three different cameras, that capture the same region of the farm but with different capture angles.
This experiment also explores the influence of image quality on our system, since the video qualities from the 3 sets are also different.

For the training images in each video set, we fine-tune three detection systems, $S1$, $S2$, and $S3$, based on each individual corresponding datasets, respectively.
An extra system $S\_all$ is trained on all the training frames together.
In the testing phase, each trained system is applied to the images from the three sets separately.
We also test each system on all testing images together for an overall comparison.
All training and testing data are separated regardless of their dataset, and no images used for both training and testing.
In total, there are 4 trained models testing on 4 groups of test sets, which forms 16 training/testing pairs.
For each pair, we measure the final system output using supervised metrics: body F1 score and leg-hoof F1 score.
Table \ref{tab:exp2} shows the comparison results. 

\begin{table}[]
    \centering
    \caption{System performance comparison on different video sets. The bold numbers show the best performance of each column.} 
    \begin{tabular}{|c||c|c|c|c||c|c|c|c|}
    \hline
    \multirow{2}{*}{\begin{tabular}[c]{@{}c@{}}Trained\\ system \end{tabular}} & \multicolumn{4}{|c||}{Body F1 score on} & \multicolumn{4}{c|}{Leg-hoof F1 score on} \\ \cline{2-9} & Set1     & Set2    & Set3    & All    & Set1      & Set2     & Set3     & All     \\ \hline
            $S1$ & 0.80 & 0.42 & 0.51 & 0.64 & 0.61 & 0.18 & 0.35 & 0.46\\ \hline
            $S2$ & 0.72 & \textbf{0.65} & 0.58 & 0.65 & 0.16 & 0.59 & 0.33 & 0.26\\ \hline
            $S3$ & \textbf{0.82} & 0.56 & 0.59 & 0.69 & 0.61 & 0.52 & \textbf{0.56} & 0.58\\ \hline
            $S\_all$ & \textbf{0.82} & 0.64 & \textbf{0.61} & \textbf{0.71} & \textbf{0.62} & \textbf{0.65} & \textbf{0.56} & \textbf{0.59}\\ \hline
    \end{tabular}
    \label{tab:exp2}
\end{table}

In Table \ref{tab:exp2}, each row represents a system trained from one dataset, and each column shows the system performance on one corresponding test set.
Comparing the four systems, it can be observed that $S\_all$ achieves similar and slightly better performance than the others, and this merged system even works better than when each individual system is both trained and tested on its own videos (diagonal values). 
This demonstrates that adding training data from other similar video sets helps to improve the detection performance. 

\begin{figure}
    \begin{subfigure}[b]{.49\linewidth}
        \includegraphics[width=\linewidth]{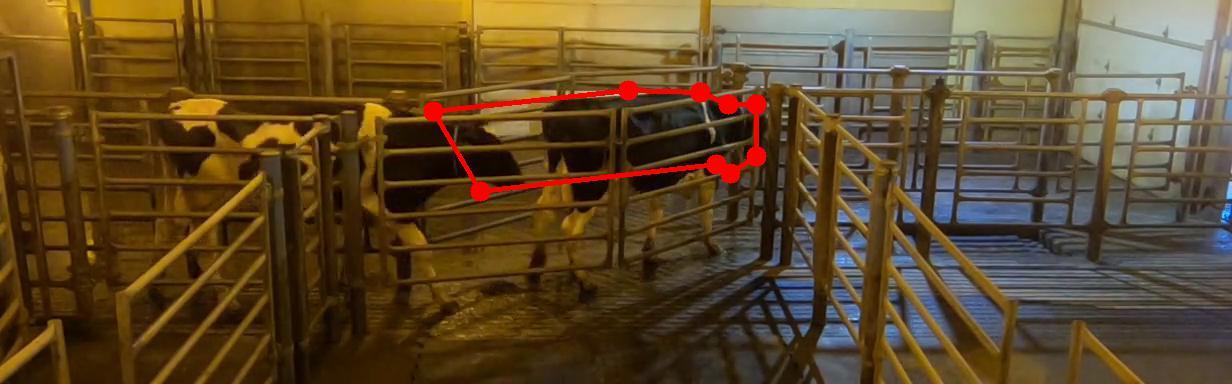}
        \caption{Output from $S1$}
        \label{fig:exp2_s1}
    \end{subfigure}
    \begin{subfigure}[b]{.49\linewidth}
        \includegraphics[width=\linewidth]{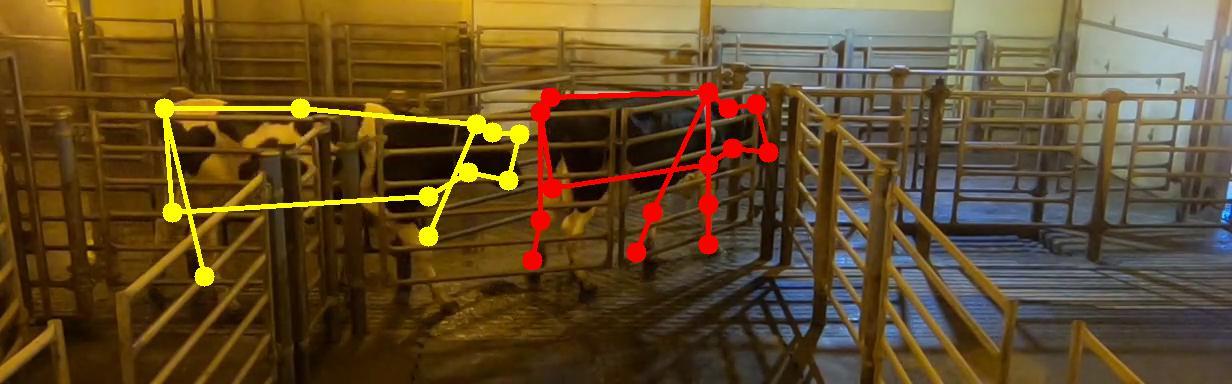}
        \caption{Output from $S2$}
        \label{fig:exp2_s2}
    \end{subfigure}
    \begin{subfigure}[b]{.49\linewidth}
        \includegraphics[width=\linewidth]{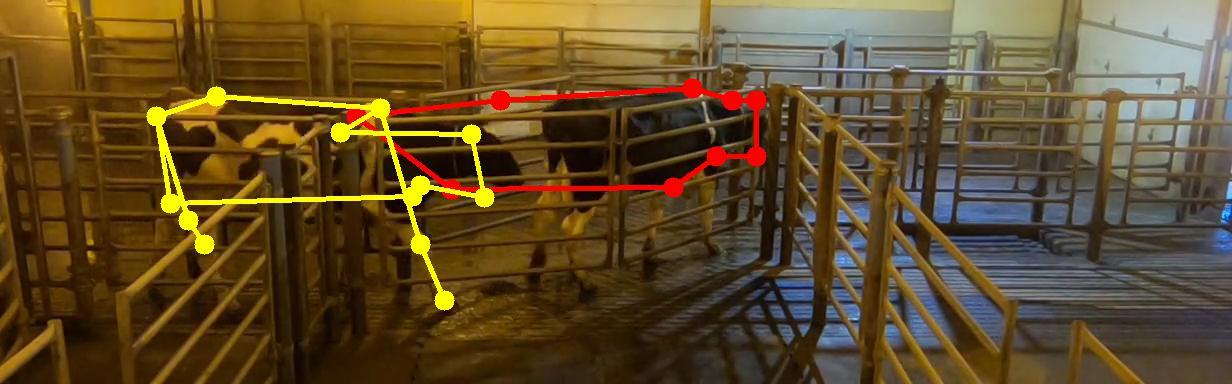}
        \caption{Output from $S3$}
        \label{fig:exp2_s3}
    \end{subfigure}
    \begin{subfigure}[b]{.49\linewidth}
        \includegraphics[width=\linewidth]{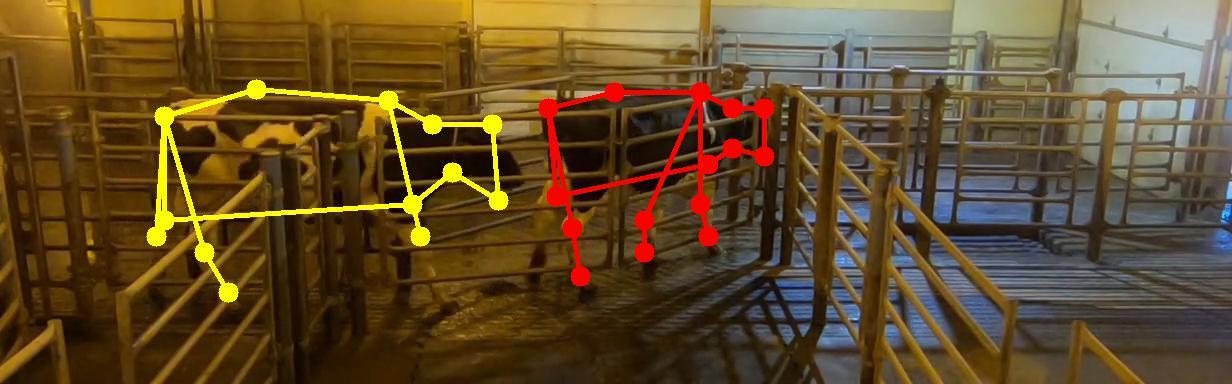}
        \caption{Output from $S\_all$}
        \label{fig:exp2_s_all}
    \end{subfigure}
    \caption{The detection comparison between systems trained on different video sets. This example image is from Set 2.}
    \label{fig:exp2}
\end{figure}

The results in Table \ref{tab:exp2} also allow us to examine the performance of the method when the input videos have different qualities. 
While both Set 1 and Set 2 have low quality, the images in Set 2 (see Figure \ref{fig:exp2}) have a small spatial resolution while the images in Set 1 (see Figure \ref{fig:exp1}) are blurry with poor illumination.
Therefore, the results of system $S1$ on Set 2 and of system $S2$ on Set 1 images are poor, especially for the leg and hoof regions.
However, system $S3$, which is trained on high quality images, provides better results on both these two datasets.
This demonstrates that using higher quality images or increasing the variation of training data can improve system performance.



A final observation from the table is that the body region F1 scores are more stable across different systems than the leg-hoof F1 scores, due to the fact that the post-processing module that only operates on the body region.
The spatial and temporal prediction in the post-processing model improve the estimation of missing and incorrectly detected points, which compensates for poor CNN performance.
Since the legs and hooves are estimated directly from the CNN outputs, the performance variation is primarily due to the variation of training data.


In addition to numerical comparison, in Figure \ref{fig:exp2} we also present some visual results from all 4 trained systems applied to a test image from Set 2 that contains two cows.
Comparing the outputs, system $S1$ fails to detect two cow objects and $S3$ is confused with some body parts between the two cow objects.
However, system $S2$ and $S\_all$ both detect two cow objects and present an accurate cow shape, because these two systems are both trained with data from Set 2.
But the merged result from $S\_all$ is more accurate on some body parts, for example the points on each cow's back, because of the additional training data involved.
However, for the leg and hoof region, none of the systems detect all the points, due to the difficulty of observing them and the lack of post-processing process.

\subsection{Segmentation methods comparison} \label{sec:experiment_mask}

This experiment compares the detection performance between our system and other popular object detection methods.
Recall that the motivation for our system is not only to segment the spatial location of the cow, but also to detect critical keypoints about its body parts. 
Therefore, ideally comparison methods should also target these two goals.
However, as mentioned in Section \ref{sec:previous_work}, most previous keypoint detection methods focus on human objects and incorporate knowledge about human body parts, and it is difficult to adapt them to cow bodies for a comparison.
On the other hand, there are many popular object detection methods which can be fine-tuned to segment cows, and these make for an effective comparison.
In this experiment, we compare the cow object detection performance between our system and other three popular pixel-wise object detection methods: One Shot Video Object Segmentation (OSVOS) \cite{caelles2017one}, DeepLab \cite{chen2017deeplab}, and Mask R-CNN \cite{he2017mask}.

\begin{figure*}
  \begin{subfigure}{\linewidth}
  \includegraphics[width=.16\linewidth]{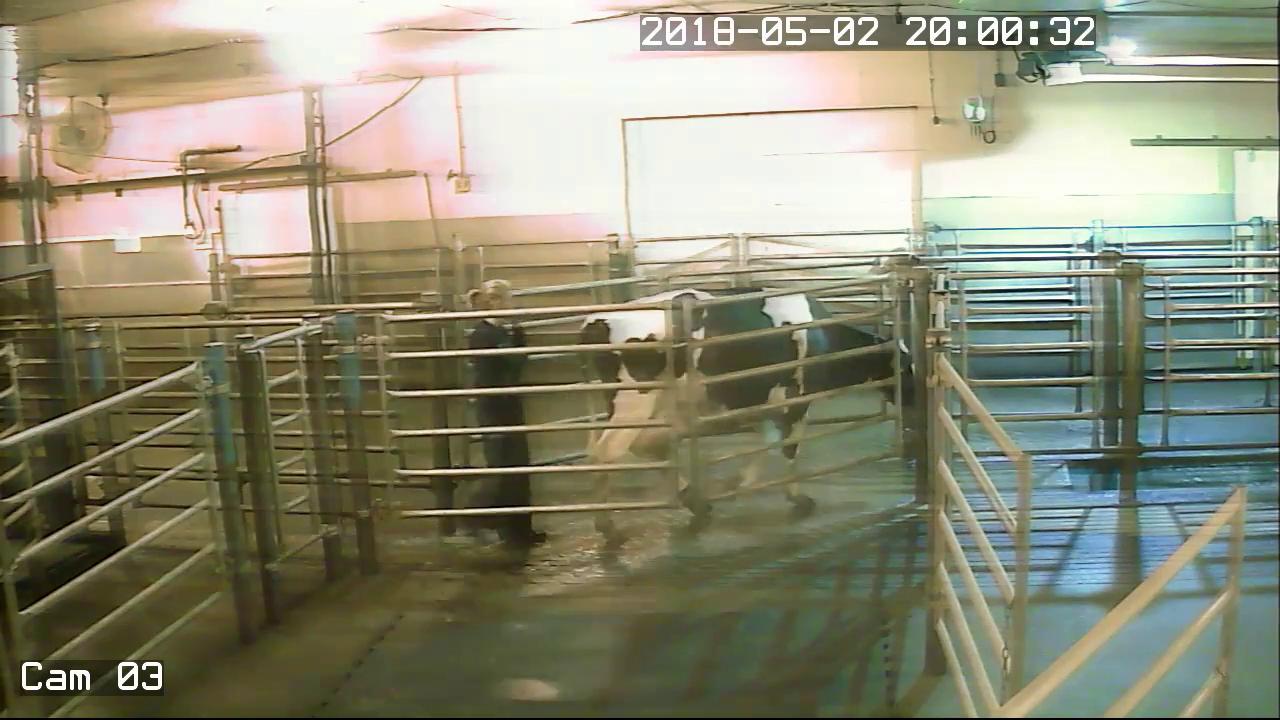}\hfill
  \includegraphics[width=.16\linewidth]{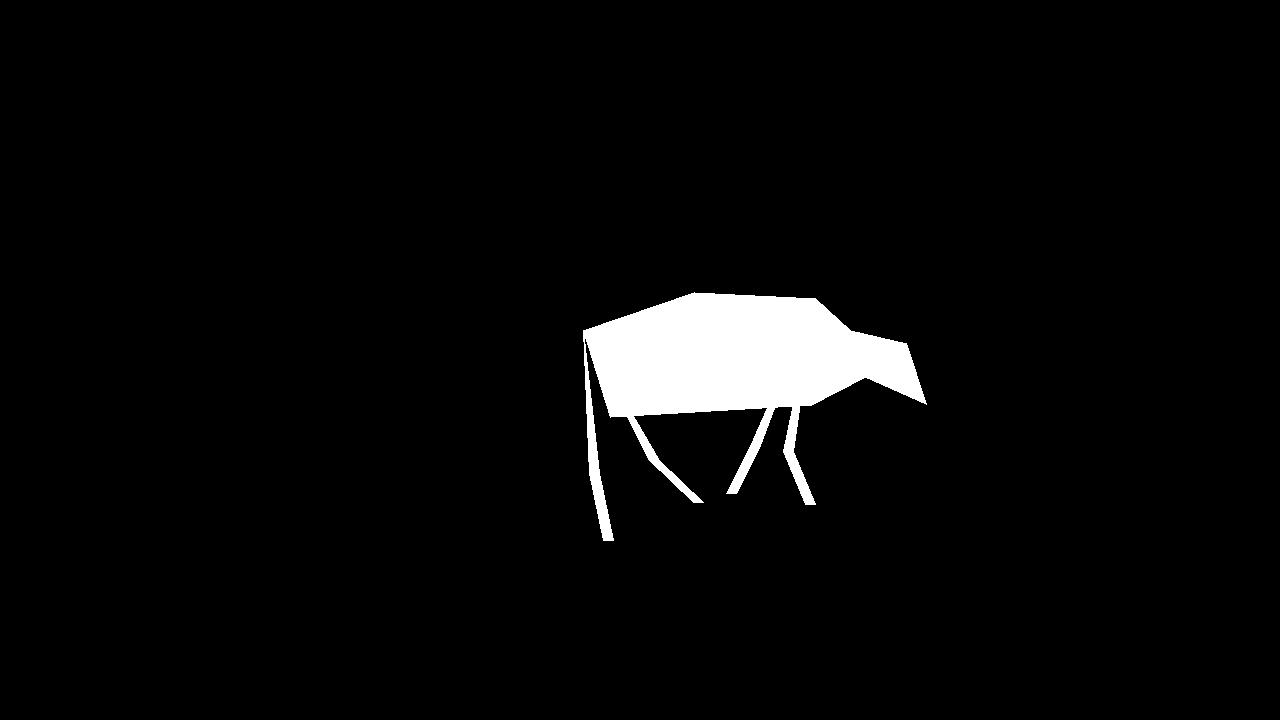}\hfill
  \includegraphics[width=.16\linewidth]{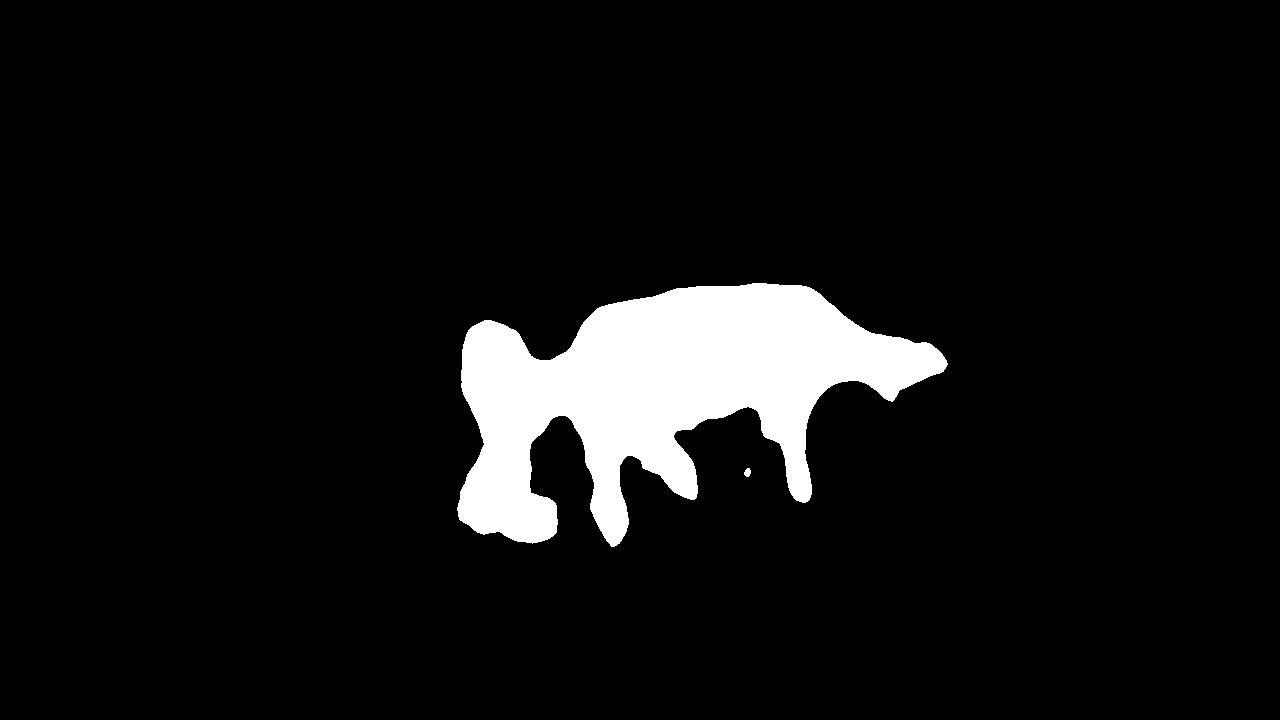}\hfill
  \includegraphics[width=.16\linewidth]{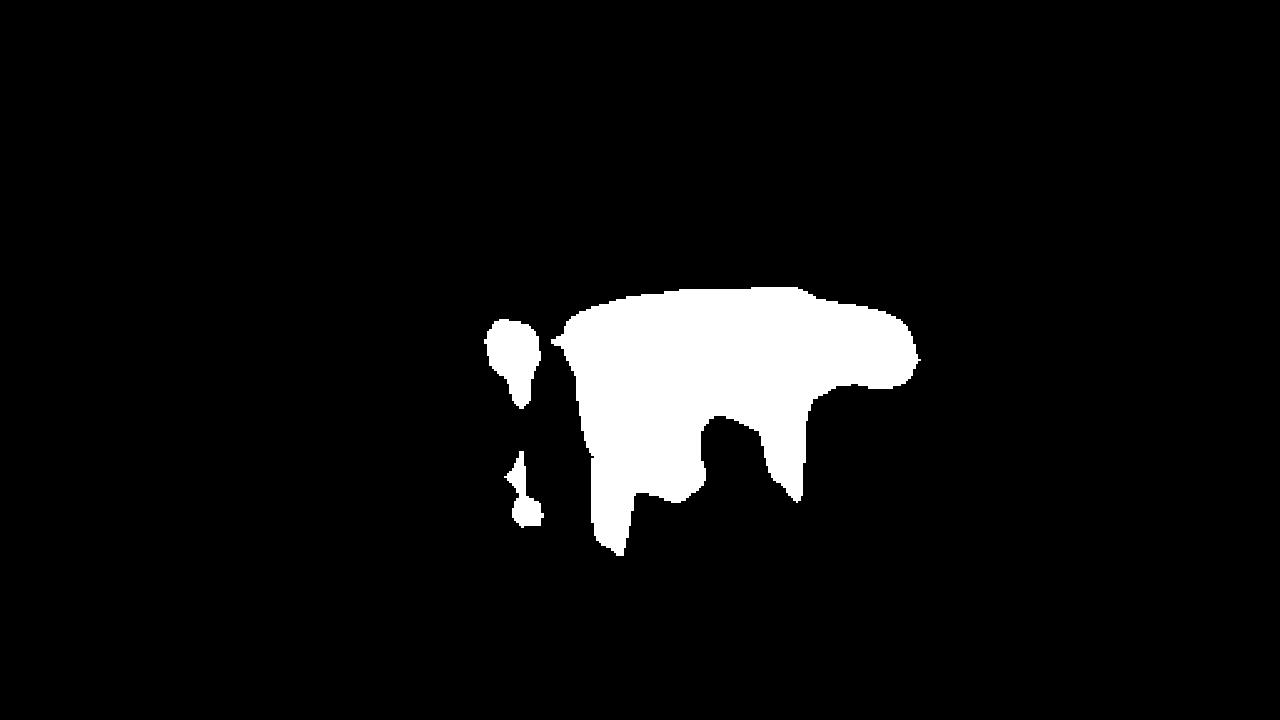}\hfill
  \includegraphics[width=.16\linewidth]{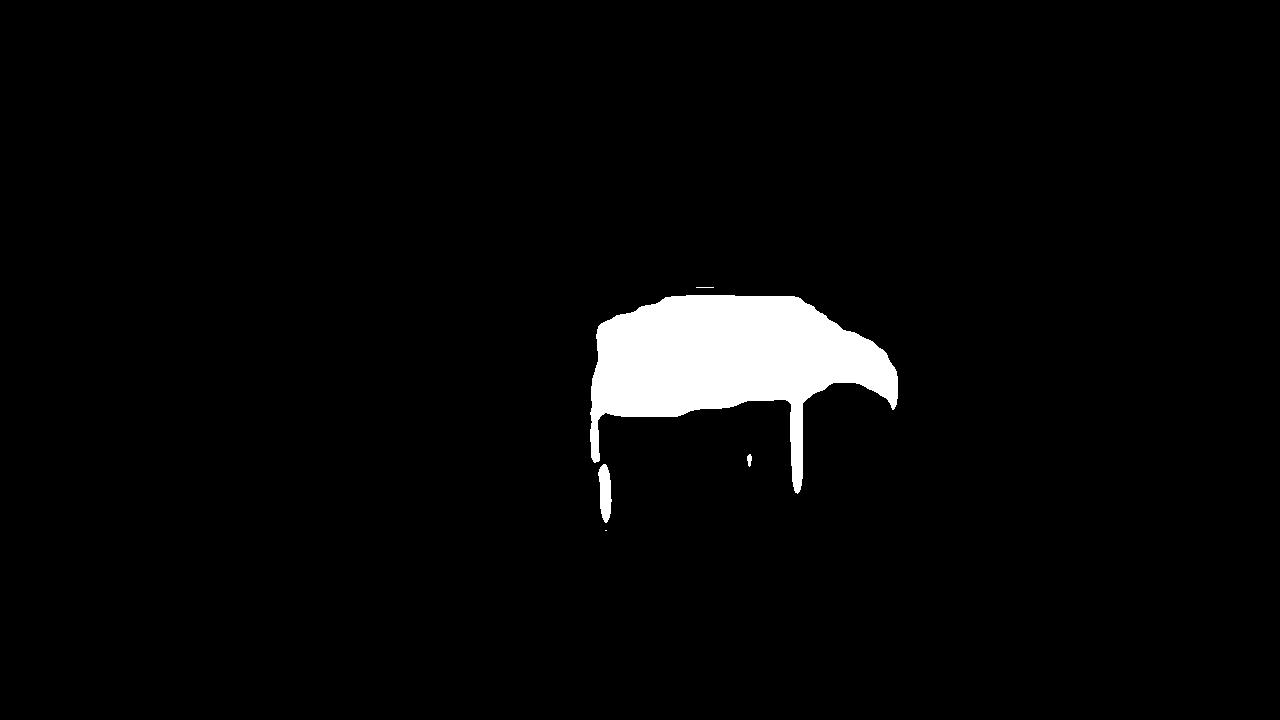}\hfill
  \includegraphics[width=.16\linewidth]{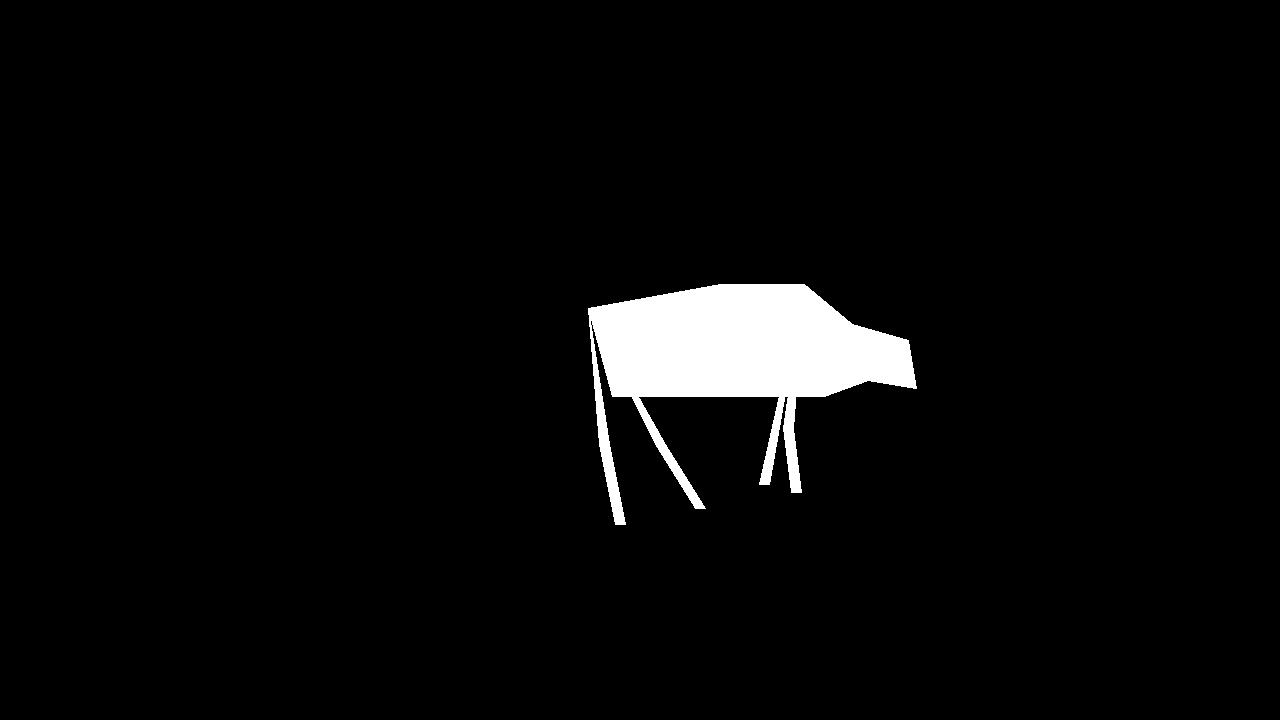}\hfill
  \caption{}
  \end{subfigure}\par\medskip
  \begin{subfigure}{\linewidth}
  \includegraphics[width=.16\linewidth]{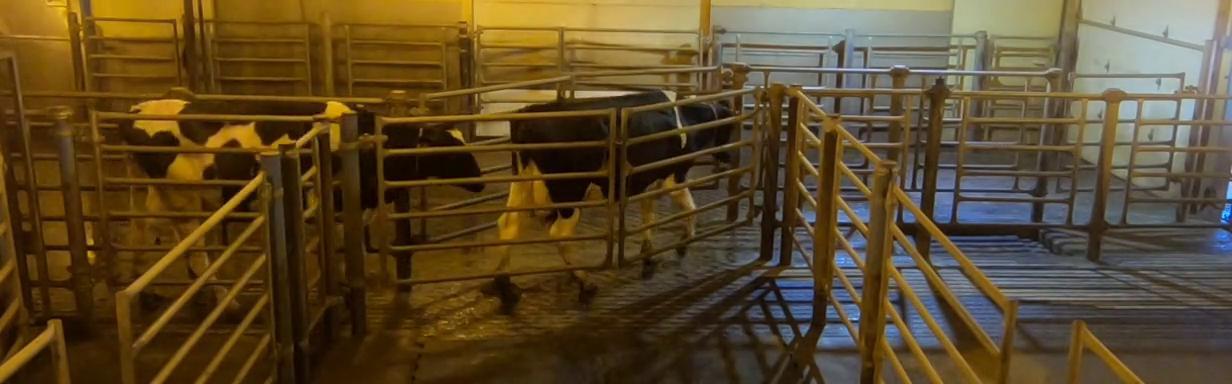}\hfill
  \includegraphics[width=.16\linewidth]{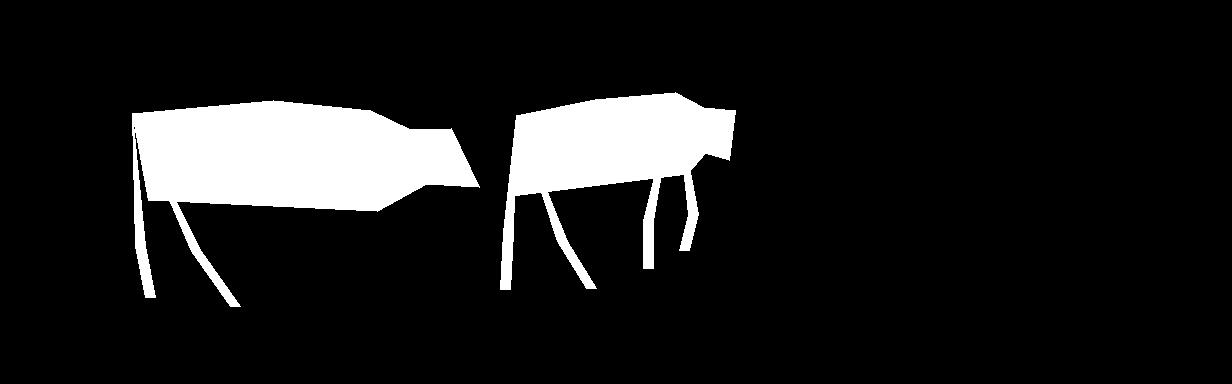}\hfill
  \includegraphics[width=.16\linewidth]{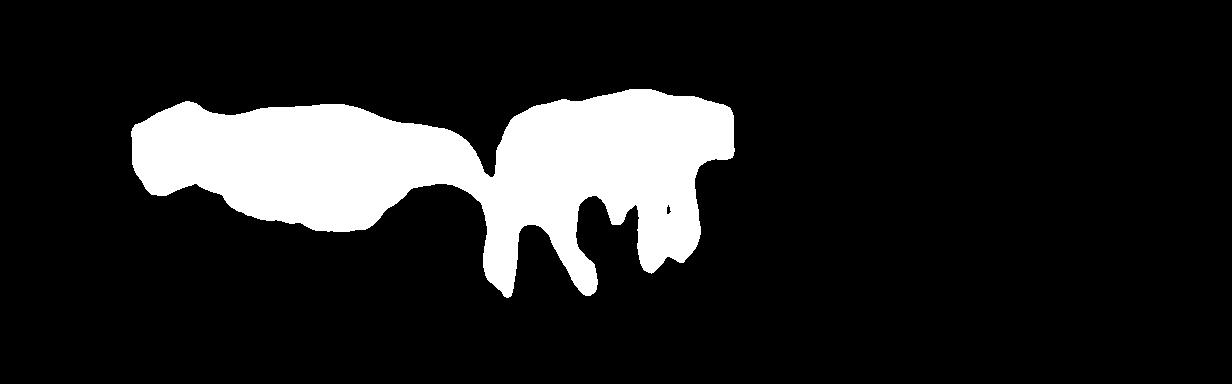}\hfill
  \includegraphics[width=.16\linewidth]{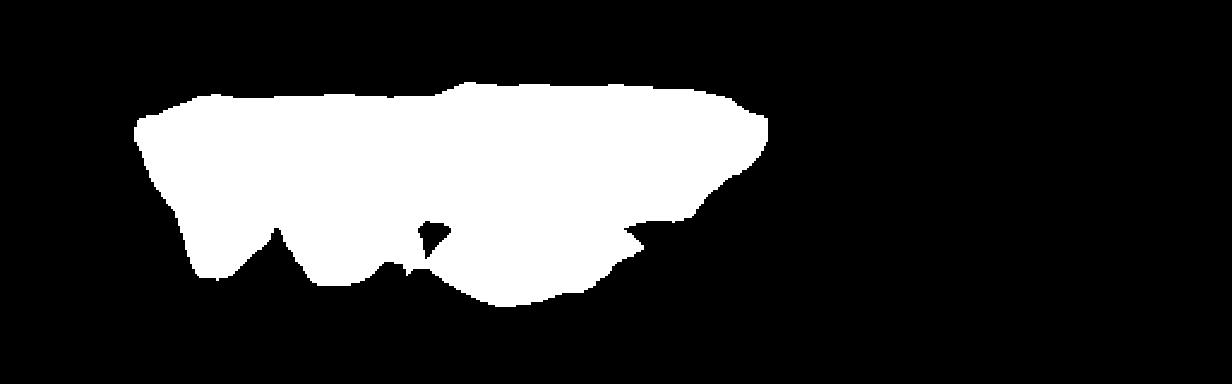}\hfill
  \includegraphics[width=.16\linewidth]{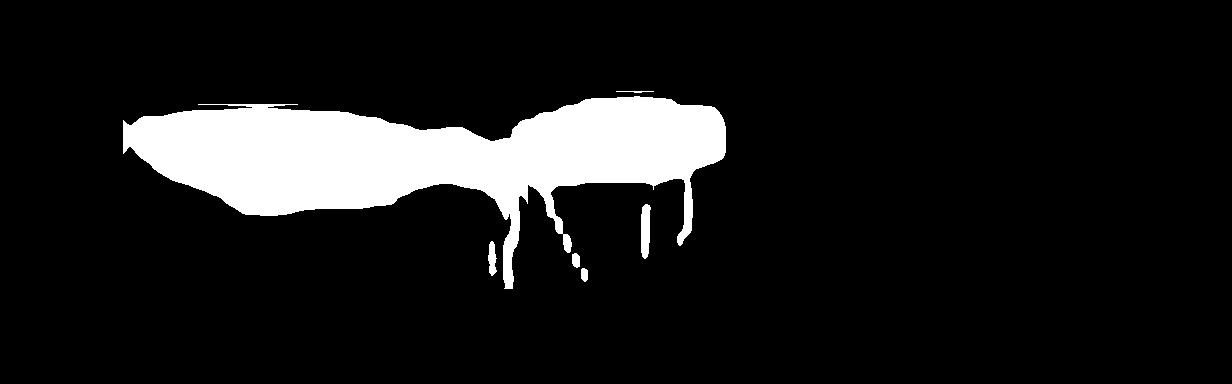}\hfill
  \includegraphics[width=.16\linewidth]{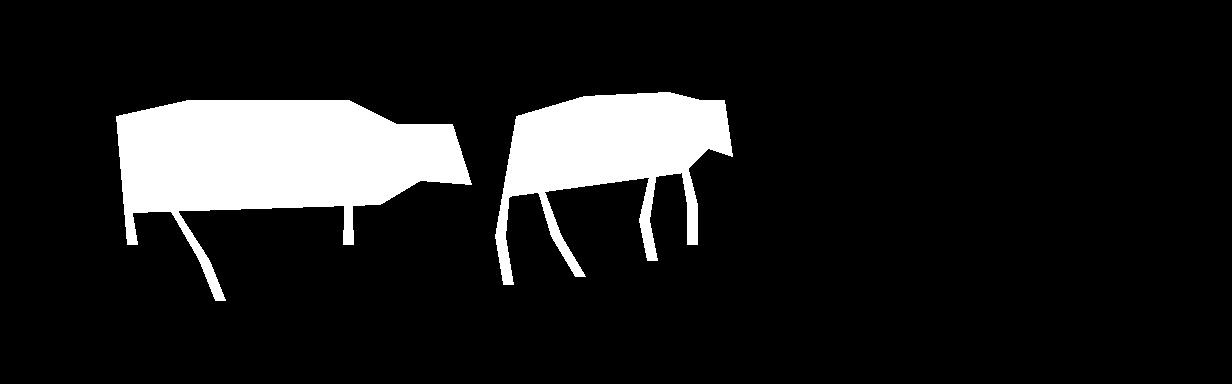}\hfill
  \caption{}
  \end{subfigure}
  \begin{subfigure}{\linewidth}
  \includegraphics[width=.16\linewidth]{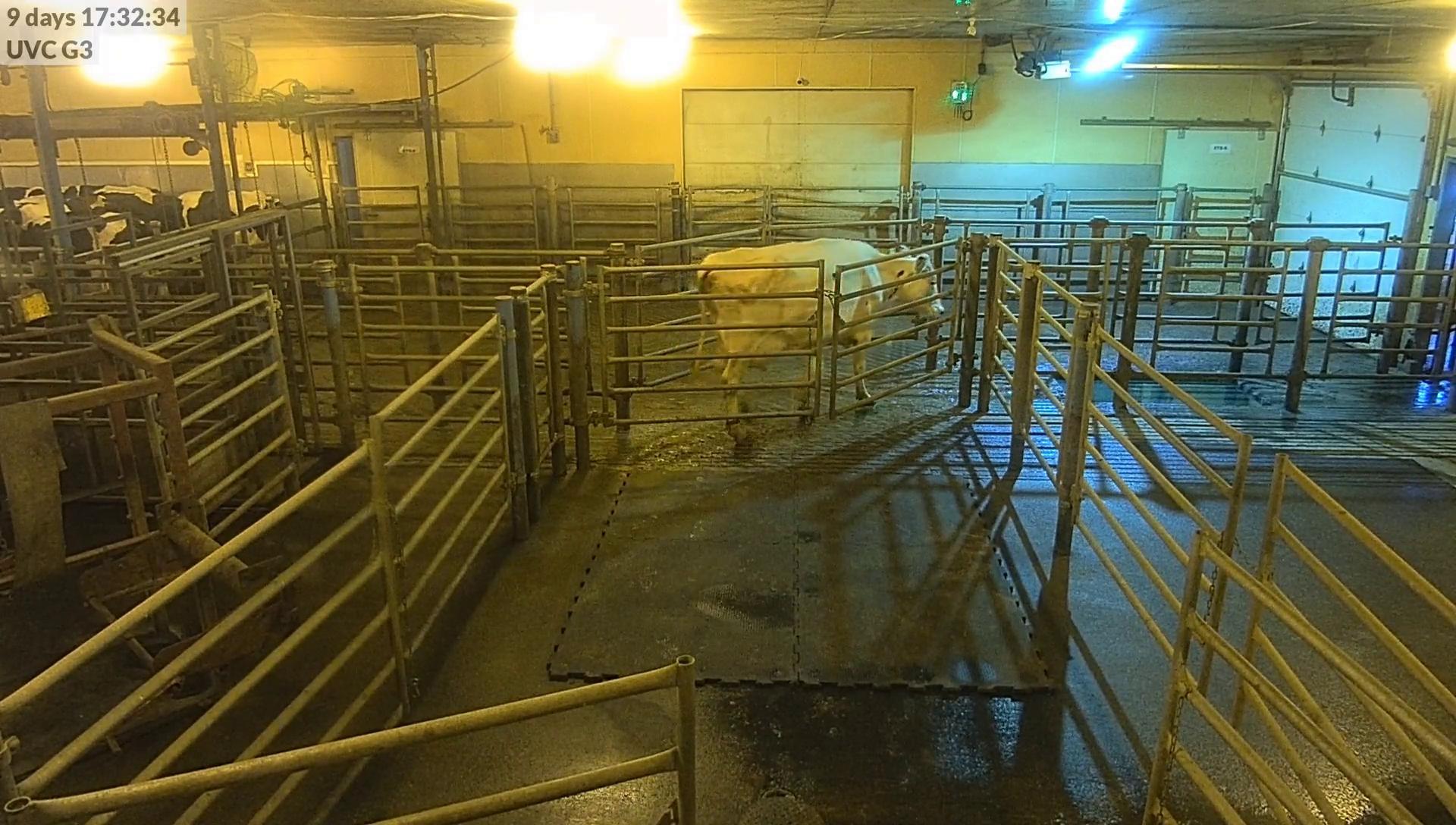}\hfill
  \includegraphics[width=.16\linewidth]{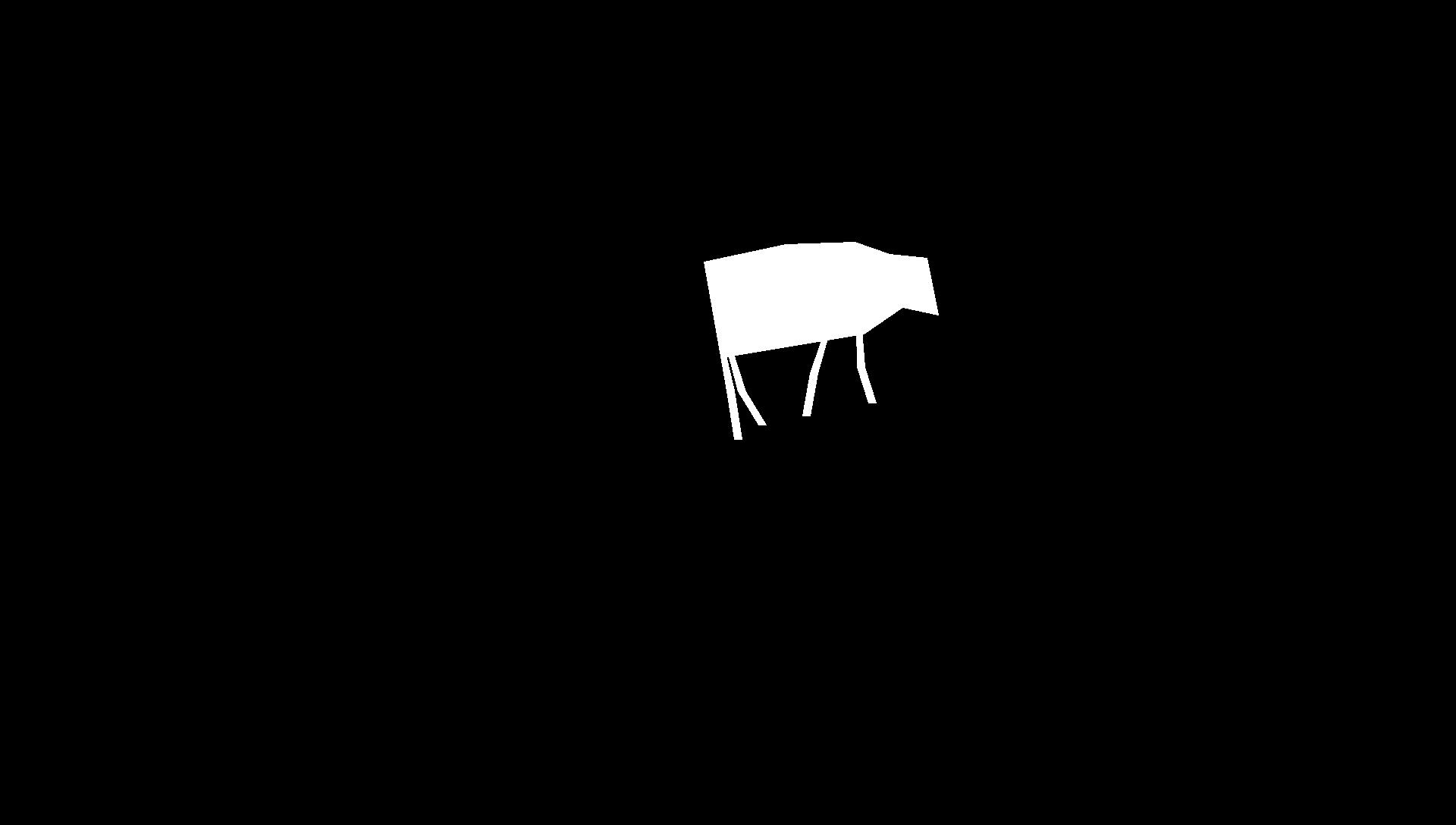}\hfill
  \includegraphics[width=.16\linewidth]{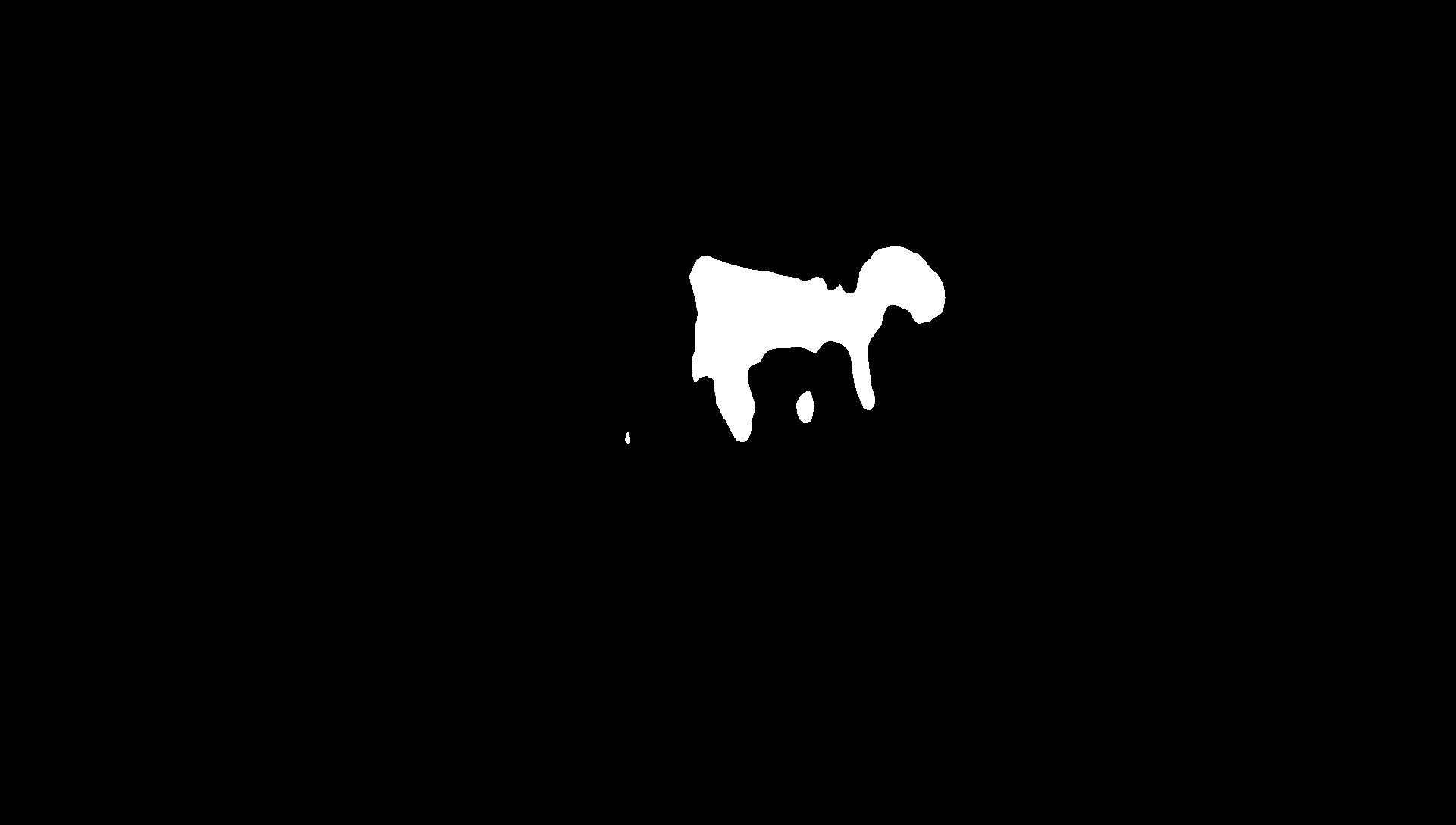}\hfill
  \includegraphics[width=.16\linewidth]{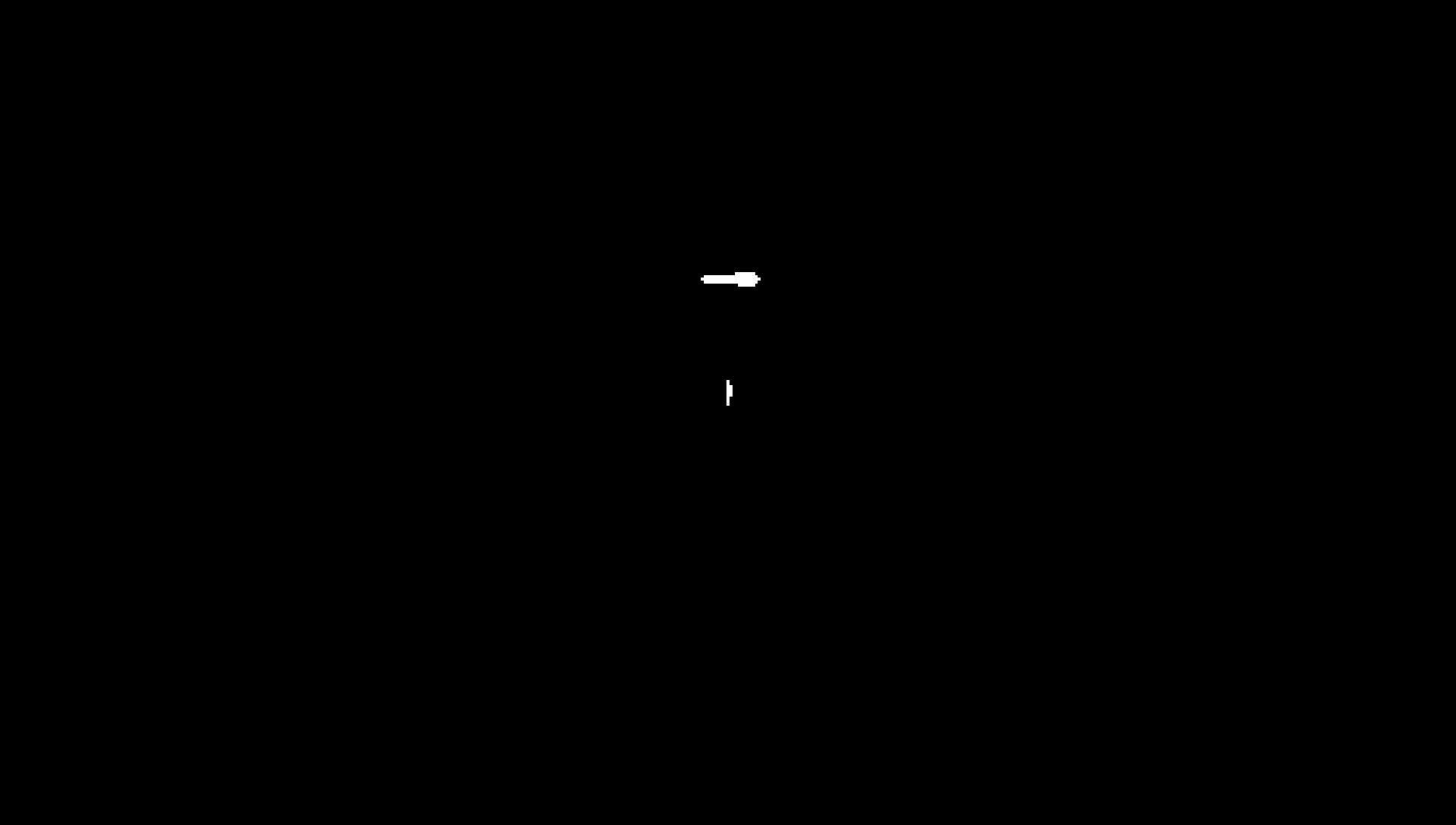}\hfill
  \includegraphics[width=.16\linewidth]{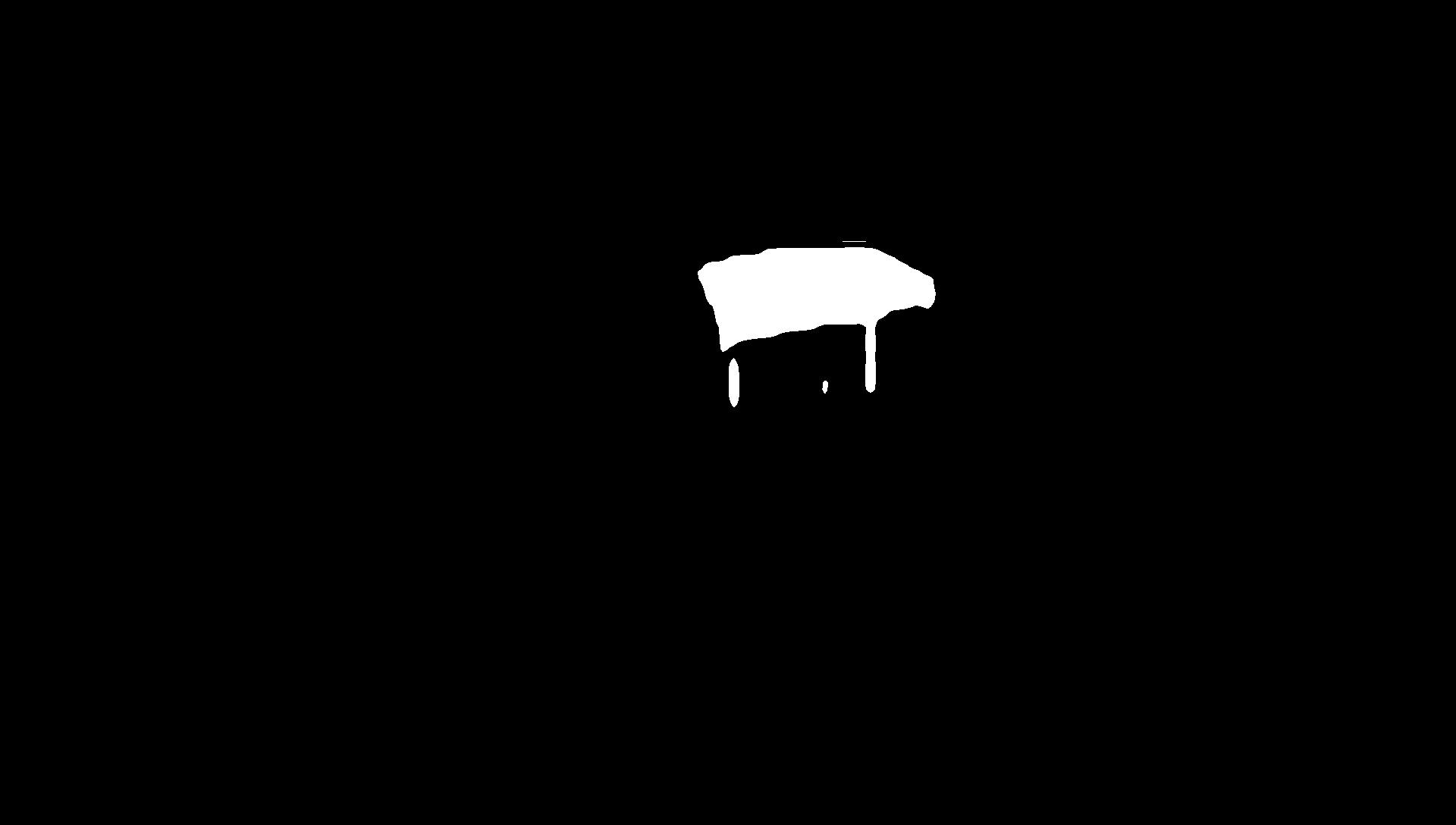}\hfill
  \includegraphics[width=.16\linewidth]{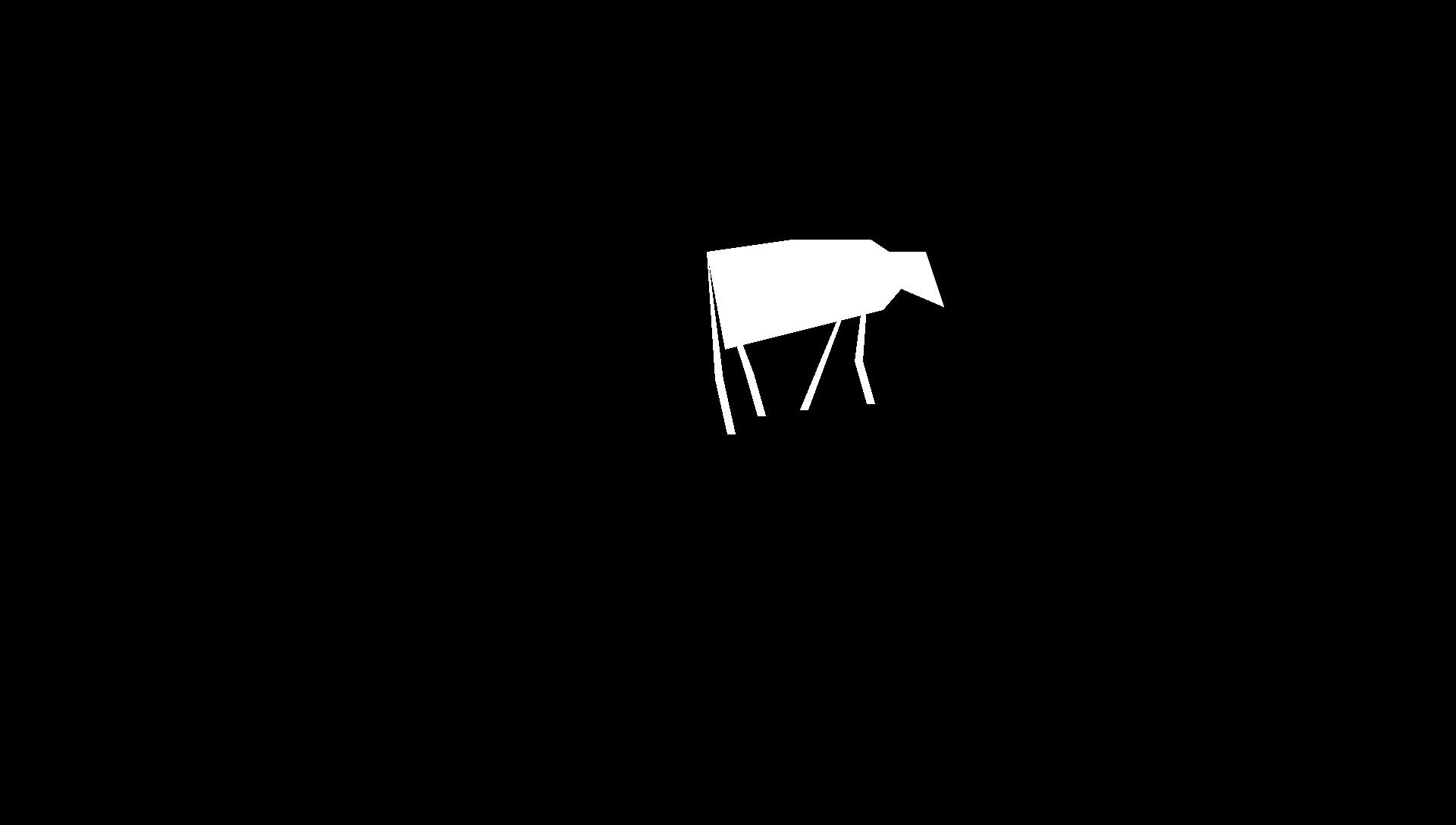}\hfill
  \caption{}
  \end{subfigure}\par\medskip
  \begin{subfigure}{\linewidth}
  \includegraphics[width=.16\linewidth]{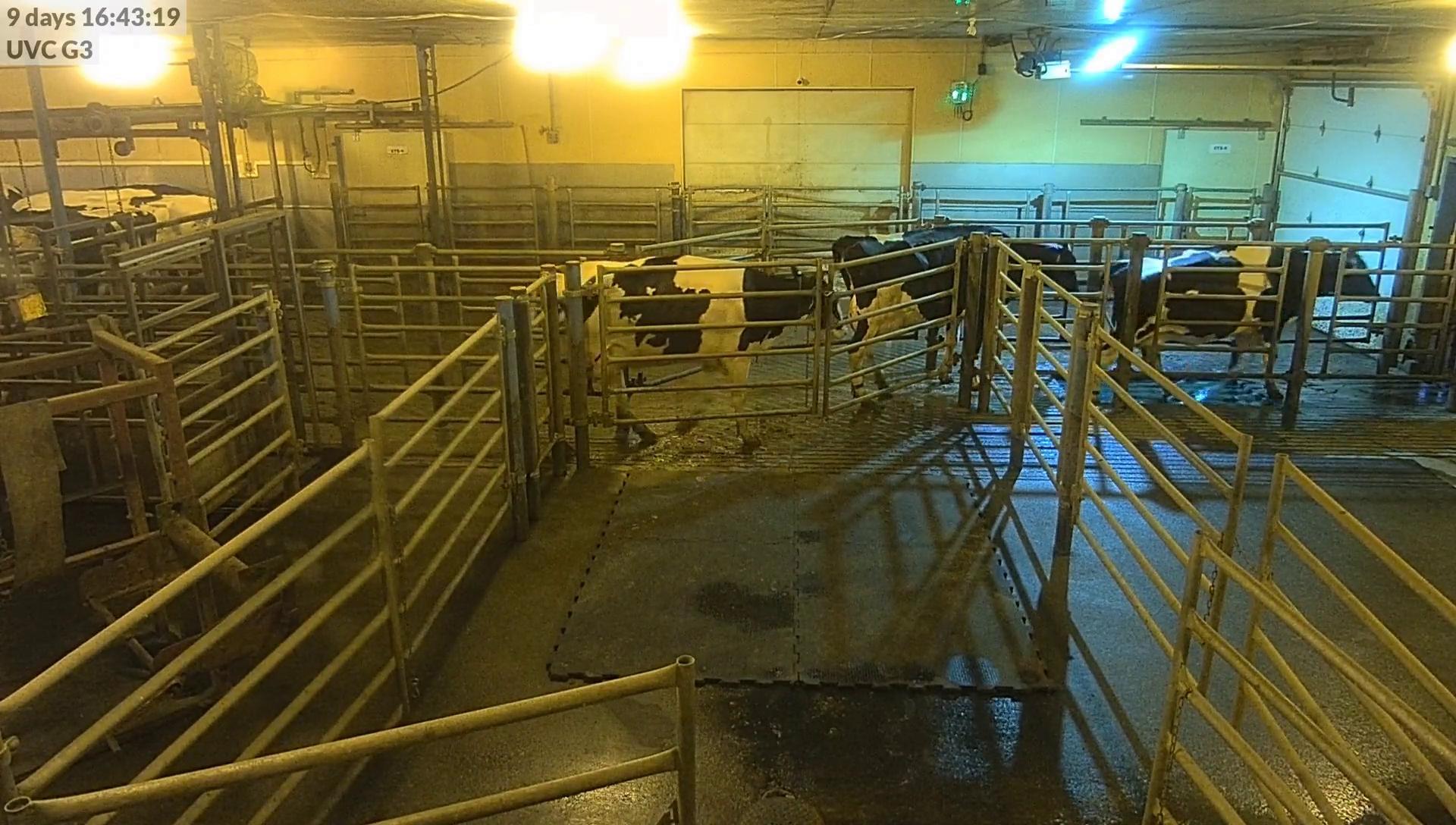}\hfill
  \includegraphics[width=.16\linewidth]{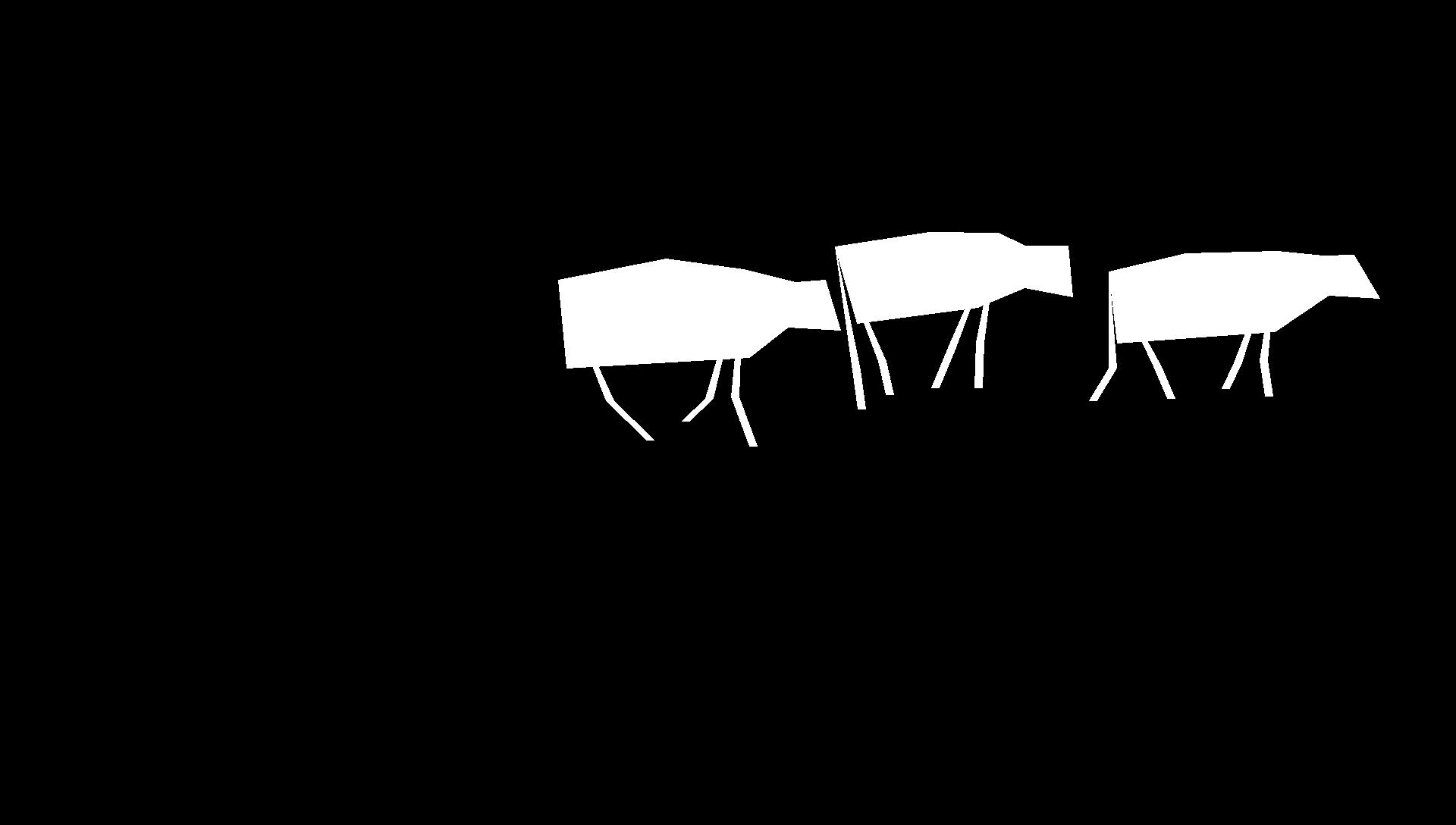}\hfill
  \includegraphics[width=.16\linewidth]{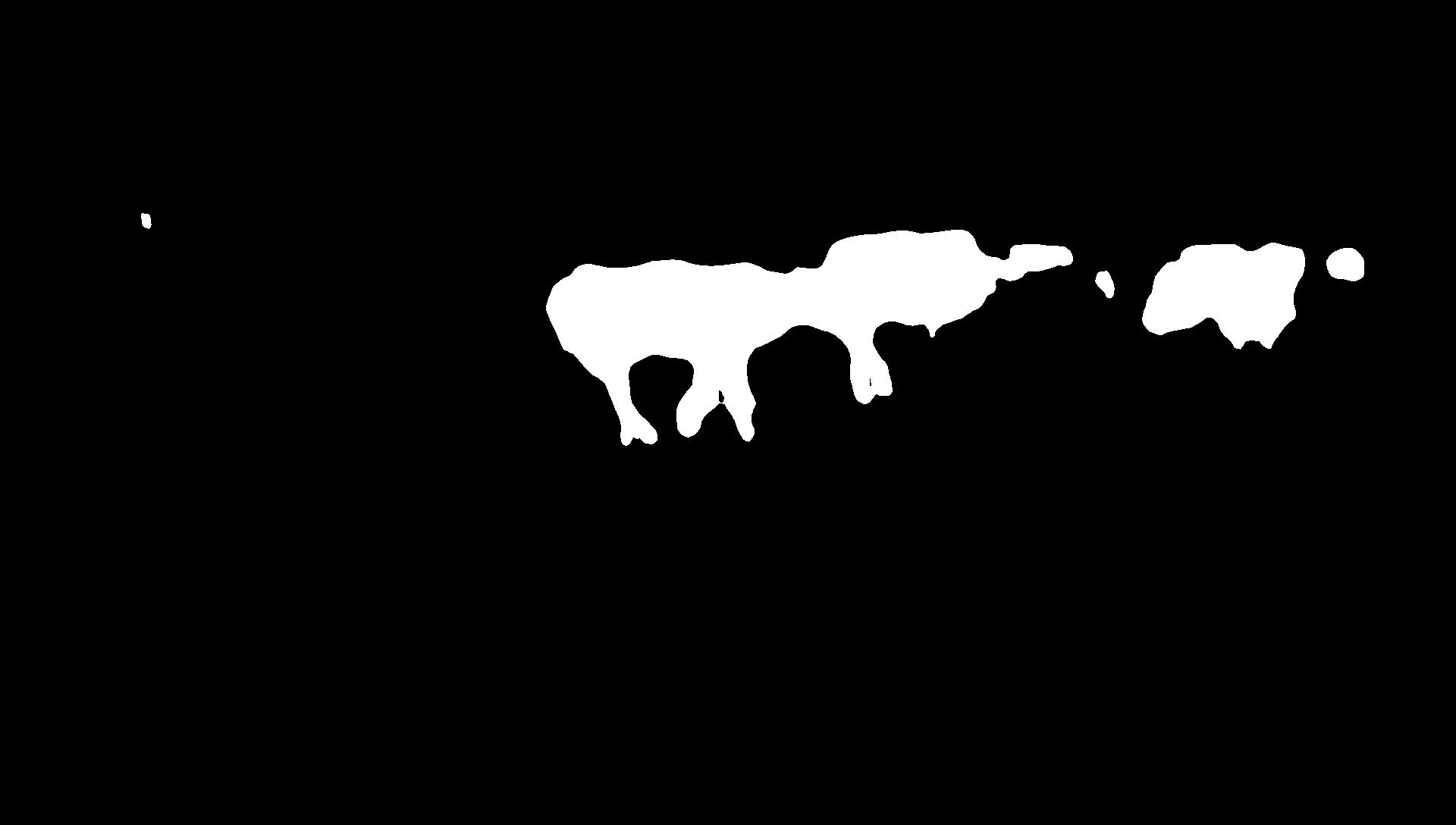}\hfill
  \includegraphics[width=.16\linewidth]{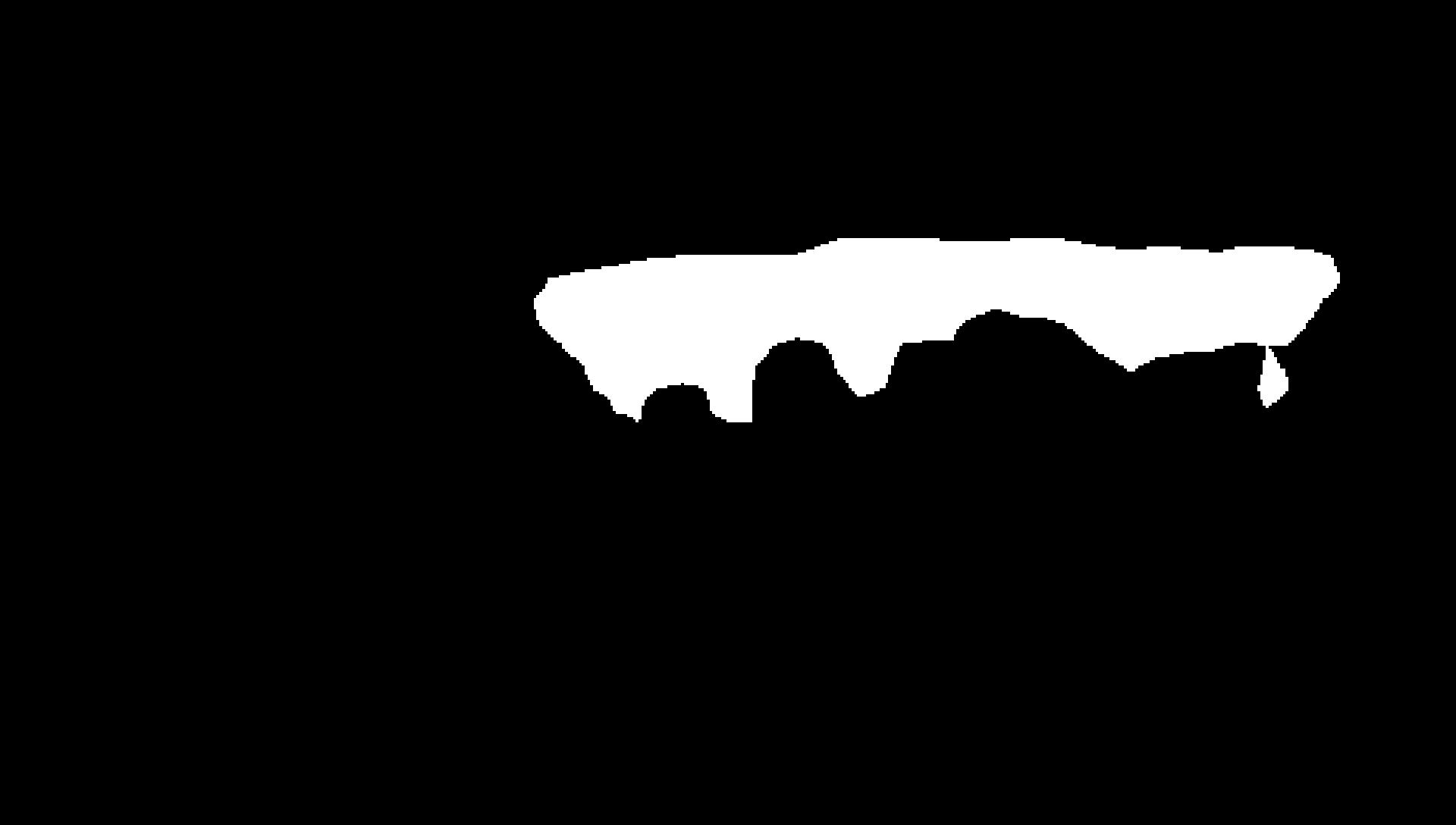}\hfill
  \includegraphics[width=.16\linewidth]{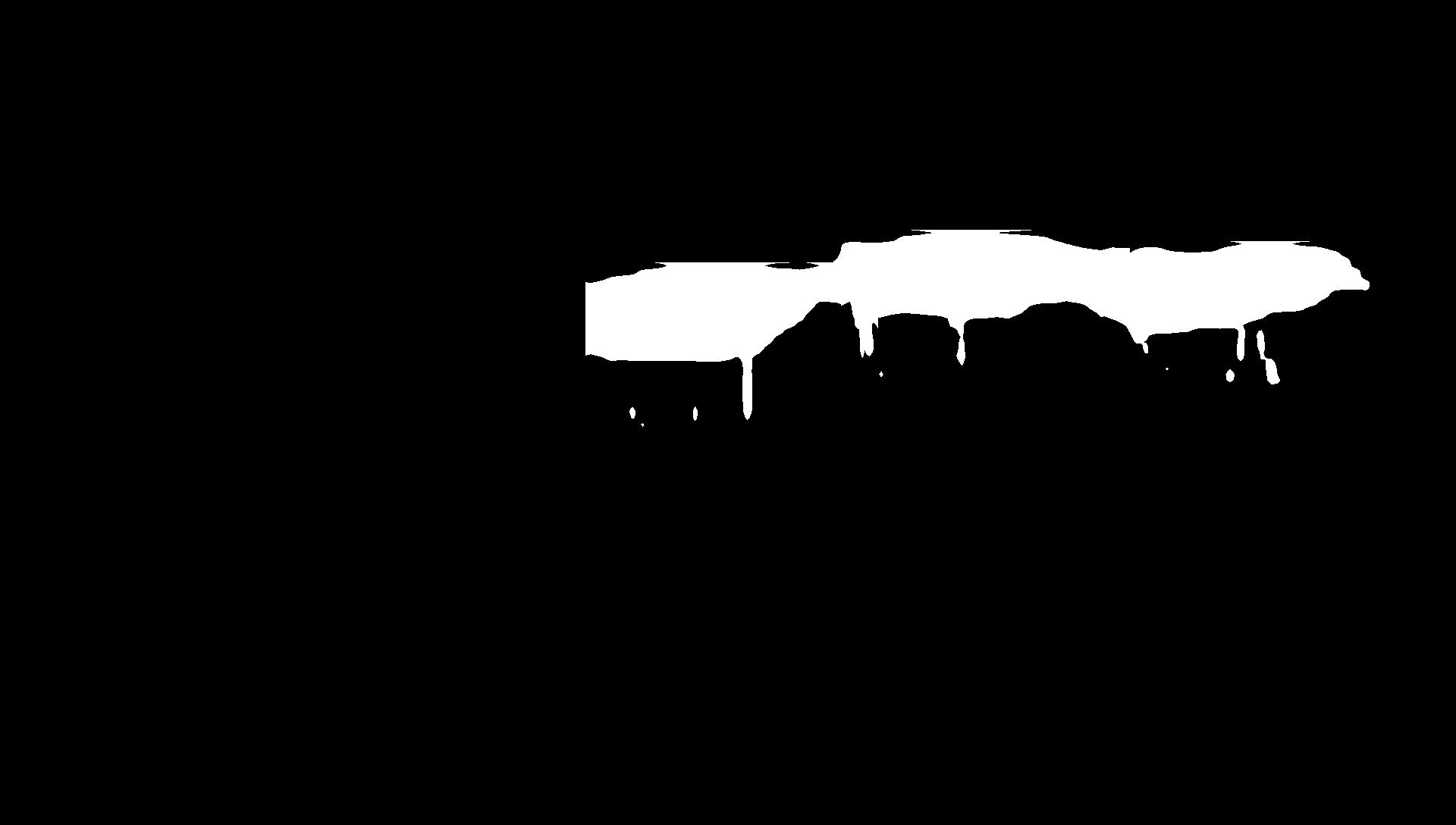}\hfill
  \includegraphics[width=.16\linewidth]{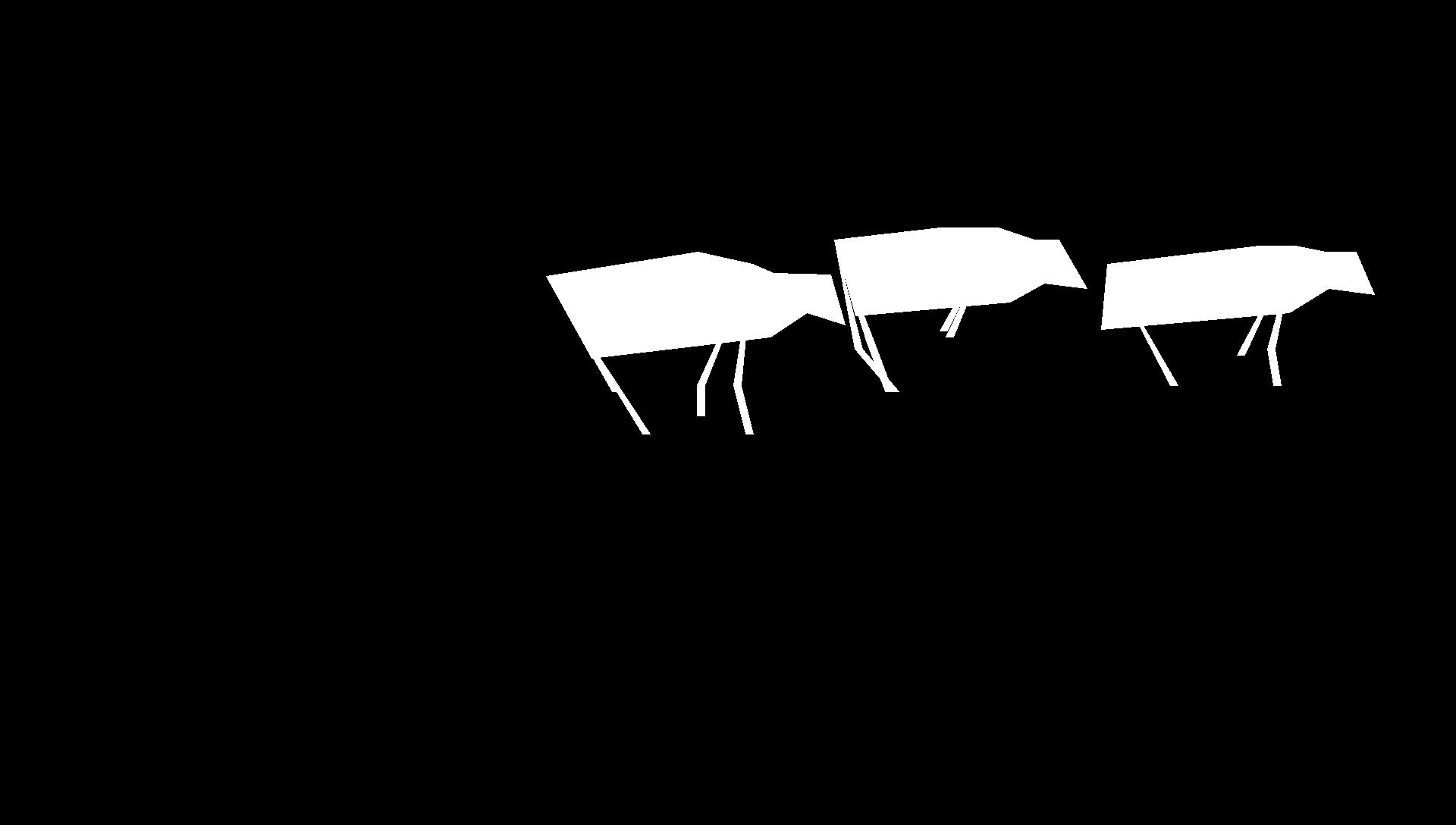}\hfill
  \caption{}
  \end{subfigure}
  \caption{Results using different detection methods. From left to right: original image, ground truth, OSVOS \cite{caelles2017one}, DeepLab\cite{chen2017deeplab}, Mask R-CNN\cite{he2017mask}, and ours. Example (a) and (b) are from Set 1 and Set 2, respectively; example (c) and (d) are both from Set 3.}
  \label{fig:exp3}
\end{figure*}

To create a performance comparison that does not disadvantage the object detection methods, we convert the output of our structural model into a binary cow mask, with two steps.
First, all keypoints from the upper body region are connected to form a closed area representing the cow body.
Second, every leg-hoof limb is expanded from a line into a polygon with a horizontal width of 20 pixels, as shown in the second column of Figure \ref{fig:exp3}.
This expansion process is applied to both the ground-truth labels and the detection results.
The newly expanded ground-truth masks are then used to fine-tune the object detection methods, as well as to compute performance metrics.
Still the point-to-mask conversion is not perfect.
Notice the approximated masks cannot exactly cover the cow object from the original image; see for example the inaccurate edges of the cow body and the straight legs.

We use all the training and testing data from the three video sets in this experiment.
In total, there are 240 single-cow frames for training and 1255 images for testing.
Each of the three comparison methods are fine-tuned with the approximate cow masks, with different implementation details.
For OSVOS \cite{caelles2017one}, we use the parent network pre-trained on the DAVIS 2016 \cite{perazzi2016benchmark} dataset and fine-tune it with our data.
The output results are binarized using Otsu \cite{liao2001fast} threshold.
For DeepLab \cite{chen2017deeplab}, we use the pre-trained network from the COCO dataset \cite{lin2014microsoft}, and we modify the last layer to produce two classes: cows and background.
The fine-tuning process is applied only on the last atrous spatial pyramid pooling layers with binary entropy loss. 
For Mask R-CNN \cite{he2017mask}, we use the network pre-trained on the COCO dataset and fine-tune its region proposal network and feature pyramid network. 
The classifier outputs are also adjusted to the two classes of cows and background.

Figure \ref{fig:exp3} shows some visual examples of the detection results.
From left to right are the original image, ground truth, and the results from OSVOS, DeepLab, Mask R-CNN, and our system, respectively.
Each row shows an example which is selected from a different test set.
Example (a) includes a human wearing black clothes who is walking right behind the cow.
This confuses OSVOS which considers it to be part of the moving foreground object.
Example (c) shows a special case which contains a pure white cow, and this color is not present in the training data.
The DeepLab method completely misses the cow, because it directly extracts information from the color image and this rare color has not been seen before.
The OSVOS method detects part of this cow using motion information, but Mask R-CNN works well because its region proposal network determines there is an object candidate and segments the cow object correctly.

Examples (b) and (d) contain multiple cows, and each method does detect multiple cow objects.
However, the three masked-based methods merge all detected cow objects together because the objects are close to each other, and we need further effort to count the number of cows or to extract other detailed information.
But our result provides a clear delineation between the cow objects, due to the use of the structural model.
Another observation about these two examples is that the cow positions in these two images are different.
Some cows are in the middle with fewer fences and others are on either the left or right side with denser fences blocking the view.
Every method can detect the middle cow, but the cows on the sides are more challenging to detect due to the obstacles. 
We further analyze the influence of fences in later paragraphs.

\begin{table}[]
    \centering
    \caption{Comparison of methods on different test sets.}
    \begin{tabular}{|c|c|c|c|c|}
    \hline
     & Set 1 & Set 2 & Set 3 & All \\ \hline
    OSVOS \cite{perazzi2016benchmark} & 0.571 & 0.580 & 0.570 & 0.571 \\ \hline
    DeepLab \cite{chen2017deeplab} & 0.655 & 0.513 & 0.577 & 0.610 \\ \hline
    Mask R-CNN \cite{he2017mask} & 0.735 & \textbf{0.692} & 0.630 & 0.682 \\ \hline
    ours & \textbf{0.750} & 0.668 & \textbf{0.662} & \textbf{0.703} \\ \hline
    \end{tabular}
    \label{tab:exp3_1}
\end{table}

\begin{table}[]
    \centering
    \caption{Comparison of methods on subsets of Set 3. \textit{Middle} means the cow is in the image center which has fewer obstacles, while \textit{Side} means the cows are on the two sides with denser fences.}
    \begin{tabular}{|c|c|c|c|c|}
    \hline
     & Middle & Side & Single-cow & Multiple-cows \\ \hline
    OSVOS \cite{perazzi2016benchmark} & 0.672 & 0.589 & 0.650 & 0.547 \\ \hline
    DeepLab \cite{chen2017deeplab} & 0.644 & 0.537 & 0.616 & 0.518 \\ \hline
    Mask R-CNN \cite{he2017mask} & \textbf{0.749} & 0.574 & 0.703 & 0.520 \\ \hline
    ours & 0.734 & \textbf{0.645} & \textbf{0.711} & \textbf{0.587} \\ \hline
    \end{tabular}
    \label{tab:exp3_2}
\end{table}

Numerical comparison results among the methods are also reported using the F1 scores of the IOU between the detection results and the ground-truth masks.
The measures are reported based on every test set separately in Table \ref{tab:exp3_1}, and on distinct subsets of Set 3 in Table \ref{tab:exp3_2}.

From Table \ref{tab:exp3_1}, it can be observed that our method achieves the highest accuracy for most sets, although its performance relative to the fine-tuned Mask R-CNN is similar.
There are three factors which may influence these scores.
First, when comparing the masks using IOU, we use a merged mask containing both the cow body and leg regions.
Since the body region occupies a larger area of the ground-truth mask, the IOU score can still be high even if the legs are miss-detected. 
Second, because the masks for our method and the ground truth are both converted from keypoints, it is highly sensitive to the positions of the keypoints, especially for the narrow leg regions. 
Small position shifts can lead to a large change to the converted mask, which will influence the IOU score.
Third, when our system does not detect a leg or hoof point, the mask will be empty in this region. 
This will also decrease the IOU of our system. 
Nonetheless, our system performs well in comparison.

As mentioned above, a main consideration of our system is to obtain acceptable performance even when there are multiple cows, and when there are obstacles like fences.
We use Set 3 videos to further explore the influence of these issues, to eliminate any performance variations due to video quality.
As Figure \ref{fig:exp3} shows, Set 3 images have a wider view of the walking path, and cows in the center have fewer fences while cows on the left or right sides are blocked with denser fences.
So we separate the testing frames from Set 3 into four subsets: cows in the middle, cows on either side, single-cow frames, and multiple-cow frames.
Among the four subsets, images with cows in the middle and with a single cow set will be easier than images from the other two subsets.
The qualitative comparison F1 scores of these subsets are shown in Table \ref{tab:exp3_2}.
From the table, Mask R-CNN has better performance on the easier test case when the cows are blocked by fewer fences.
But for difficult test sets like denser obstacles, our proposed system works better.
The OSVOS method also performs well when there are more obstacles because this method only considers the foreground and background, which allows it to separate the stationary fences from the moving cows.

In general, compared to the other three mask-based object detection methods, our proposed system has three advantages.
First, based on the keypoints detection, our method can correctly detect the cow structure even when the cows are behind the fences or there are humans nearby.
Second, when there are multiple cow objects, this system can explicitly isolate each cow even when they are close to each other.
Third, it can detect cows with color patterns that do not exist in the training data through the use of frame difference images.
However, our system also has two limitations. 
First, the cow structural model completely depends on the accuracy of the body parts, and one inaccurate detection can cause large errors for the body contour and influence the overall spatial location. 
Second, the prediction system in our method is based on the keypoint constraints from the cow structural model, which is fixed after the training process.
If there are not enough cow body parts detected, the prediction system still forces the results to conform to a particular shape, which could cause incorrect results.

\section{Conclusion} \label{sec:conclusion}

In this work, we design a practical system to detect the structural information for cows recorded in video.
We use keypoints to form a cow structural model, which represents both the cow's overall spatial location and the positions of its specific body parts, such as the joints from the leg and hoof.
The proposed detection system applies two CNNs to extract the keypoints from raw images, and a post-processing model is developed to select individual points and convert them into cow structural models.
This system can detect and track multiple cow objects at the same time, and it also works with different quality videos which are captured on commercial farms during normal operation.

\bibliographystyle{IEEEtran}
\bibliography{references.bib}

\end{document}